\documentclass[aps,preprint,preprintnumbers,amsmath,amssymb,amsfonts,groupedaddress,superscriptaddress,showpacs,showkeys,floatfix]{revtex4-1}

\usepackage{graphicx}
\usepackage{subfigure}
\usepackage{bm}% bold math
\usepackage{color}
\usepackage{array}
\usepackage{caption}
\usepackage{caption}
\newcommand{\bmath}{\begin{mathletters}}
\newcommand{\emath}{\end{mathletters}}
\newcommand{\be}{\begin{eqnarray}}
\newcommand{\ee}{\end{eqnarray}}
\newcommand{\ba}{\begin{array}}
\newcommand{\ea}{\end{array}}
\newcommand{\la}{\langle}

\newcommand{\ra}{\rangle}

\newcommand{\no}{\nonumber}

\newcommand{\pr}{\prime}

\newcommand{\calD} {\mathcal D}

\newcommand{\calM} {\mathcal M}

\newcommand{\calR} {\mathcal R}

\newcommand{\rmd} {\mathrm{d}}

\newcommand{\Tr} {\mathrm{Tr}}

\begin{document}
\title{Simulating a Topological Transition in a Superconducting Phase Qubit by Fast Adiabatic Trajectories}
\author{Tenghui Wang}
\affiliation{Physics Department, Zhejiang University, Hangzhou, 310027, China}
\author{Zhenxing Zhang}
\affiliation{Physics Department, Zhejiang University, Hangzhou, 310027, China}
\author{Liang Xiang}
\affiliation{Physics Department, Zhejiang University, Hangzhou, 310027, China}
%\author{Jiadong Yao}
%\affiliation{Physics Department, Zhejiang University, Hangzhou, 310027, China}
\author{Zhihao Gong}
\affiliation{Physics Department, Zhejiang University, Hangzhou, 310027, China}
\author{Jianlan Wu }
\affiliation{Physics Department, Zhejiang University, Hangzhou, 310027, China}
\author{Yi Yin \footnote{Correspondence and requests for materials should be addressed to Y.Y. (email: yiyin@zju.edu.cn) or to J.L.W.
(email: jianlanwu@zju.edu.cn).}}
\affiliation{Physics Department, Zhejiang University, Hangzhou, 310027, China}
\affiliation{Collaborative Innovation Center of Advanced Microstructures, Nanjing, 210093, China}

\begin{abstract}
The significance of topological phases has been widely recognized in the
community of condensed matter physics. The well controllable quantum systems
provide an artificial platform to probe and engineer various topological
phases. The adiabatic trajectory of a quantum state describes the
change of the bulk Bloch eigenstates with the momentum, and this adiabatic simulation method is however practically
limited due to quantum dissipation. Here we apply the `shortcut to adiabaticity' (STA)
protocol to realize fast adiabatic evolutions in the system of a superconducting phase qubit.
The resulting fast adiabatic trajectories illustrate
the change of the bulk Bloch eigenstates in the Su-Schrieffer-Heeger (SSH) model. A sharp transition
is experimentally determined for the topological invariant of a winding number.
Our experiment helps identify the topological Chern number of a two-dimensional
toy model, suggesting the applicability of the fast adiabatic simulation
method for topological systems.

\end{abstract}

\maketitle

\noindent{\bf Introduction}\\
The study of topological phases has been an emerging field in condensed matter
physics since the discovery of the integer quantum Hall effect~\cite{klitzingPRL80}.
In the traditional Landau theory of phase transition, each phase is characterized by an order parameter.
Instead, various phases in a topological material are distinguished by their different
topological invariants. In the
theory of Thouless, Kohmoto, Nightingale and den Nijs, %(TKNN),
the integer Chern number is a topological invariant to interpret a
quantized Hall conductivity of a two-dimensional (2D) electronic gas~\cite{ThoulessPRL82}.
Similar topological invariants are defined in other topological systems.
For the one-dimensional (1D) Su-Schrieffer-Heeger (SSH) model~\cite{SSHmodel79}, %shown in Fig.~\ref{fig_n00},
the topologically nontrivial phase with edge states
is characterized by a unity winding number of the bulk structure according
to the bulk-boundary correspondence~\cite{Asboth2016short}.

The rapid progress of quantum manipulation techniques
has attracted much attention of simulating topological phases using
controllable quantum systems, such as cold atoms, superconducting
qubits and nitrogen-vacancy (NV) center in diamond ~\cite{KitagawaNatCommu11,AtalaNatPhy13,JotzuNat14,AidelsburgerNatPhys15,
DucaSci15,MittalNatPhoton16,FlaschnerSci16,GoldmanNatPhys16,RoushanNat14,SchroerPRL14,Flurin2016,KongFeiPRL16}.
%Instead of creating an artificial many-body system,
%The noninteracting band structure of topological systems allows a mapping onto
%a single artificial particle~\cite{xxx}.
For the SSH model and other two-band systems,
the bulk Hamiltonian in the momentum space is equivalently described by a spin-half
particle subject to a changing magnetic field. %external field. %\color{red}{In the intuitive adiabatic method,
An adiabatic trajectory of the spin simulates the %changing
bulk Bloch eigenstates as the momentum traverses the first Brillouin zone (FBZ).
The topological invariant of a Bloch band %, such as the Chern number,
is subsequently obtained by integrating a local geometric quantity %,such as Berry curvatures,
over the closed area of the FBZ~\cite{ThoulessPRL82,Asboth2016short,HasanRMP10}.
The key of this adiabatic simulation is to realize
the adiabatic evolution of a quantum state,
which is also relevant in quantum information
and quantum computation~\cite{Farhi2000,ChuangBook}.

However, a slow adiabatic operation is practically challenging
since the surrounding environment inevitably
destroys quantum coherence at a long time scale.
%The topological transition
%was often simulated  to circumvent the difficulty of adiabaticity~\cite{xxx}.
Several strategies have been
proposed to speed-up the operation while maintaining adiabaticity~\cite{DemirplakJPCA03,Berry2009,XiChenPRL10,Masuda2009,Torrontegui2013ShortcutReview,TorosovPRL11,MartinisPRA14}.
The `shortcut to adiabaticity' (STA) protocol is
a general methodology, in which a counter-diabatic Hamiltonian %is supplemented to
cancels the non-adiabatic deflection of a quantum state~\cite{DemirplakJPCA03,Berry2009,XiChenPRL10,Masuda2009,Torrontegui2013ShortcutReview}.
%With the STA protocol, the quantum state transfer can be operated in a short time
%scale, following a designed adiabatic trajectory. %\color{red}{The high fidelity state
%manipulation with STA}
%The state transfer under
The STA protocol has been implemented in a few quantum
systems, such as cold atoms and a nitrogen-vacancy center in a
diamond~\cite{BasonNatPhys12,JFZhangPRL13,ZhouNatPhys16,AnNatCommu16}.
In a recent experiment, we applied the STA protocol to make a fast measurement
of the Berry phase in a superconducting phase qubit~\cite{ZZXPRA17}.

%\begin{figure}%[t!]
%\includegraphics[width=0.65\linewidth]{fig0_SSH.eps}
%\caption{{\bf Two topological phases of the SSH model.}
%({\bf a}) A conventional insulator from $\Omega_1>\Omega_2$. ({\bf b}) A `topological' insulator from $\Omega_1<\Omega_2$.
%Here $\Omega_1/2$ ($\Omega_2/2$) is the intracell (intercell) hopping amplitude.
%Each dashed ellipse represents a unit cell consisting
%of two sites, $A$ and $B$.
%}
%\label{fig_n00}
%\end{figure}

In this article, we simulate the topological transition of the SSH model based on
fast adiabatic trajectories of a superconducting phase qubit under the STA protocol.
%transferred from the ground to excited states.
To remove the influence of higher excited states,
the fast adiabatic state transfer is improved by the derivative removal by adiabatic gates (DRAG)
method~\cite{MotzoiPRL09,GambettaPRA11,Lucero10}.
%The two-band Hamiltonian of the
%SSH model is mimicked by the combination of a constant and rotating external fields.
To simulate the evolution of the bulk Bloch eigenstates, the fast adiabatic trajectories are
generated and measured in both real-time and virtual ways.
%two methods are proposed to generate the `adiabatic' trajectories in real-time and virtual ways.
As the intracell hopping amplitude varies, the change of the adiabatic trajectories
illustrates the transition from a topologically nontrivial to trivial phase. An integration over
the measured trajectory of the quantum state leads to a sharp change of the winding number.
Our investigation is extended to a 2D model, where the transition of
the Chern number is observed.
\\

\noindent{\bf Results} \\
\noindent {\bf Fast adiabatic state transfer following the STA protocol.}
In the rotating frame of a microwave drive  pulse, a two-level superconducting qubit
is mapped onto a spin-half particle. The Hamiltonian is written as
$H(t) = (\hbar/2)\bm B_0(t)\cdot \bm \sigma$, where $\bm \sigma=(\sigma_x, \sigma_y, \sigma_z)$ is the vector of Pauli operators
and $\bm B_0(t)$ is an effective magnetic field in the unit of angular frequency.
%This rotating frame ($\calR$-frame) rather than the laboratory frame is the starting point of describing our experiment.
A slowly-varying external field $\bm B_0(t)$ drives the spin to follow an instantaneous eigenstate of $H(t)$.
For instance, we consider a rotating field, $\bm B_0(t) = (\Omega\sin\theta(t), 0, \Omega\cos\theta(t))$,
in the $x$-$z$ plane, where $\Omega$ is the drive amplitude and $\theta(t)$ is the time-varying polar angle.
Through the evolution of the instantaneous spin-up state, a quantum state transfer from the qubit ground ($|0\rangle$)
to excited state ($|1\rangle$) is realized
when $\theta(t)$ is evolved from 0 to $\pi$.
In a simple manner, we apply a sinusoidal pulse where
the polar angle is linearly increased with time, i.e., $\theta(t)=(\pi/T_a)t$~\cite{LuPRA11}.
To satisfy the adiabatic theorem, a long operation time $T_a$ is required,
which is however difficult in our phase qubit due to relatively short relaxation time ($T_1= 310$ ns)
and pure decoherence time ($T^\ast_2= 120$ ns).

\begin{figure}[t!]
\includegraphics[width=0.75\linewidth]{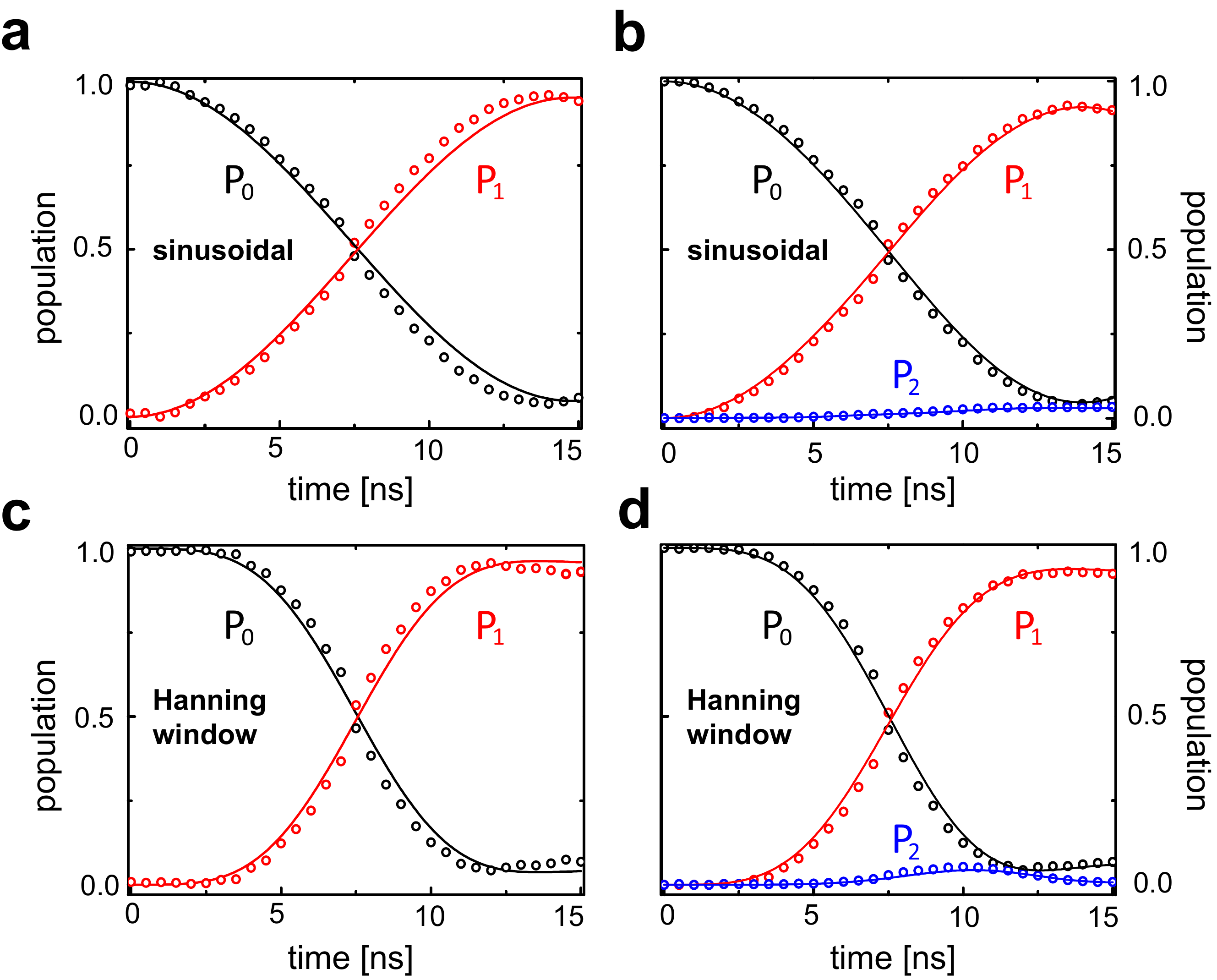}
\caption{ Population evolutions
subject to a sinusoidal STA field in the frameworks of  the ({\bf a}) two-level and ({\bf b}) three-level systems.
Population evolutions subject to a Hanning-window STA field in the frameworks of the ({\bf c}) two-level and ({\bf d}) three-level systems.
In {\bf a}-{\bf d}, the populations of the ground ($|0\rangle$), first excited ($|1\rangle$) and second excited ($|2\rangle$) states are shown in
black, red and blue colors, respectively. The symbols are the experimental measurements while the solid
lines are the corresponding Lindblad calculations.
}
\label{fig_n01}
\end{figure}

Instead, we implement the STA protocol to achieve a fast adiabatic state transfer (see Methods).
An additional counter-diabatic field,  $\bm{B}_\mathrm{cd}(t)=(0, \dot{\theta}(t),0)$ with $\dot{\theta}(t) =\pi/T_a$, is included
and the modified Hamiltonian becomes $H(t)=(\hbar/2)\bm B(t)\cdot \bm \sigma$ with $\bm B(t)=\bm B_0(t)+\bm{B}_\mathrm{cd}(t)$.
In an ideal scenario, %this counter-diabatic field
$\bm{B}_\mathrm{cd}(t)$ cancels the non-adiabatic transition
so that the spin follows exactly the same path of $\bm B_0(t)$~\cite{Berry2009}. %when subject to the STA field $\bm B(t)$~\cite{xxx}.
%In our experiment, $\bm B(t)$
%is accurately produced by our sophisticated microwave generators~\cite{xxx}.
The drive amplitude is set as $\Omega/2\pi=30$ MHz, and the operation time is $T_a=15$ ns
which is on the same time scale as a fast $\pi$-pulse.
The qubit is initially reset at the ground state. %$|0\rangle$.
%and the state transfer takes the adiabatic trajectory of $|s_\uparrow(t)\rangle$.
The STA field $\bm B(t)$ is interrupted every 0.5 ns to measure the population in the framework
of a two-level system.
As shown in Fig.~\ref{fig_n01}a, the population of the excited state is increased with time,
close to the theoretical prediction, $P_1(t)=(1/2)[1-\cos\theta(t)]$.
The final population transferred %of the excited states
is $P_1(t=T_a)=0.943$,
with a small deviation from an ideal result. %adiabatic state transfer.
The numerical calculation of the Lindblad equation
%with the relaxation time $T_1= 310$ ns and the pure decoherence time $T^\ast_2= 120$ ns
is used to inspect the influence of qubit dissipation~\cite{ChuangBook}.
In Fig.~\ref{fig_n01}a,  a small but visible difference is observed
between the experimental measurement and the Lindblad calculation, mainly in the second half of the
STA operation.
%This observation suggests an additional influence of higher excited states
%on the deviation of our fast `adiabatic' state transfer~\cite{xxx}.

Since the phase qubit arises from a multi-level anharmonic oscillator~\cite{MartinisReview},  %the framework of
a three-level  system, $\{|0\rangle, |1\rangle, |2\rangle\}$,
is employed to re-examine the state transfer process.
The Hamiltonian is changed to $H(t)=(\hbar/2)\bm B(t)\cdot \bm S+\hbar\Delta_2|2\rangle\langle 2|$.
The vector $\bm S=(S_x, S_y, S_z)$ is defined as
$S_x=\sum_{n=0}^1 \sqrt{n+1}(|n+1\rangle\langle n|+h.c.)$, $S_y=\sum_{n=0}^1 \sqrt{n+1}(i|n+1\rangle\langle n|+h.c.)$,
and $S_z = \sum_{n=0}^2 (1-2n) |n\rangle\langle n|$.
A large anharmonic frequency shift, $\Delta_2/2\pi=-200$ MHz, exists in our system.
By experimentally projecting onto the three quantum states,
the corrected population evolutions are plotted in Fig.~\ref{fig_n01}b.
A small but nonzero population $P_2(t)$ of the second excited state is increased with time,
causing a population leakage out of  %the qubit subspace,
$\{|0\rangle, |1\rangle\}$.
The actual populations transferred in the STA operation are
$P_1(t=T_a)=0.907$ and $P_2(t=T_a)=0.036$.
The necessarity of the three-level system is also confirmed by
good agreement between the Lindblad calculation and the experimental measurement. %(Fig.~\ref{fig_n01}b).

%In an approximate way,
The final population at the second excited state is approximated as
$P_2(t=T_a)\approx \dot{\theta}(t=T_a)^2/[2(\Delta_2+\Omega)^2]$ (see Methods).
To reduce its influence, we apply an constraint of $\dot{\theta}(t=T_a)=0$
to re-design the polar angle in a Hanning-window form, i.e.,  $\theta(t) = (\pi/2)[1-\cos(\pi t/T_a)]$.
The counter-diabatic field, $\bm{B}_{\rm cd}(t)=(0, \dot{\theta}(t),0)$, is modified accordingly.
The experimental population evolutions subject to the new STA field $\bm B(t)$ are plotted in Fig.~\ref{fig_n01}c,d
in the frameworks of the two-level and three-level systems, respectively.
The population at the first excited state is increased to $P_1(t=T_a)=0.923$ while
the population at the second excited state is decreased to $P_2(t=T_a)=0.009$.
Due to a more efficient control on the population leakage,
the Hanning-window pulse rather than the simpler sinusoidal pulse will be under investigation
in the rest of this paper.
\\

\begin{figure}[t!]
\includegraphics[width=0.75\linewidth]{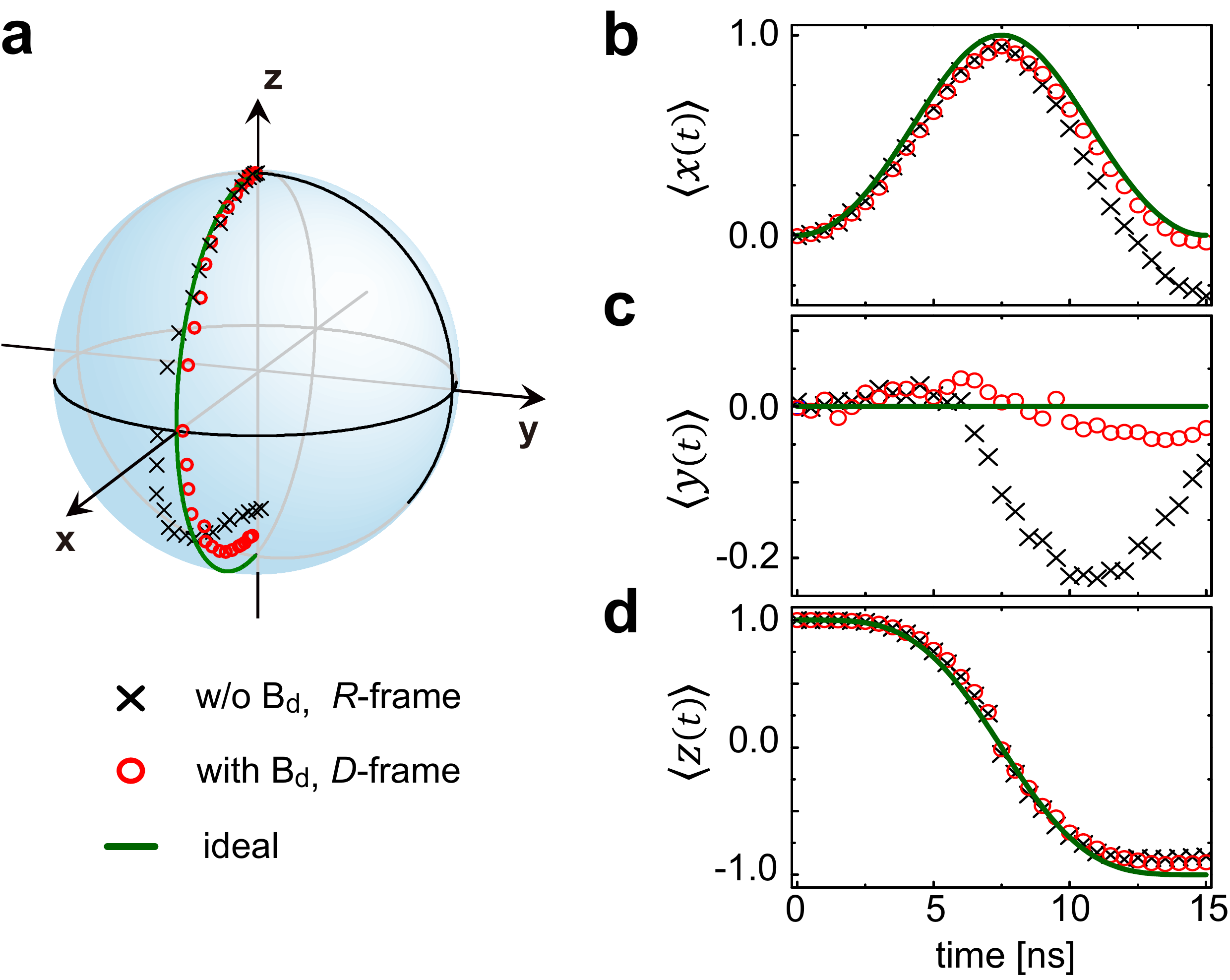}
\caption{
({\bf a}) The trajectories of the qubit vector depicted on the Bloch sphere.
The time evolutions of the three projections: ({\bf b}) $\la x(t) \ra$, ({\bf c}) $\la y(t) \ra$ and ({\bf d}) $\la z(t) \ra$.
In {\bf a}-{\bf d}, black crosses denote the experimental result without the DRAG correction and under the $\calR$-frame,
red circles denote the experimental result with the DRAG correction and under the $\calD$-frame,
and the green lines denote the theoretical prediction of an ideal adiabatic state transfer.}
\label{fig_n02}
\end{figure}

\noindent {\bf Derivative removal by adiabatic gates.}
To visualize the trajectory of the entire state transfer process, the quantum
state tomography (QST) is performed every 0.5 ns to extract the density matrix $\rho(t)$~\cite{LuceroNatPhys12}.
For convenience, the QST measurement is restricted to the qubit subspace, $\{|0\rangle, |1\rangle\}$.
The experimental Bloch vector, $\bm r(t)=(\langle x(t)\rangle, \langle y(t)\rangle, \langle z(t)\rangle)$,
is determined by the three projections, $\langle \zeta(t)\rangle = \Tr\{\rho(t)\sigma_\zeta \}$ with $\zeta=x, y, z$.
The trajectory of $\bm r(t)$ is depicted on the Bloch sphere in Fig.~\ref{fig_n02}a,
while the time evolutions of the three projections  are plotted in Fig.~\ref{fig_n02}b-d respectively.
Compared to the ideal trajectory, $\bm r(t)=(\cos\theta(t), 0, \sin\theta(t))$,
the experimental qubit vector gradually shrinks inside the Bloch sphere due to the qubit dissipation.
A severer distortion is observed in the $x$-$y$ plane,
where both $\langle x(t)\rangle$ and  $\langle y(t)\rangle$ deviate from
the ideal path (black crosses versus green lines in Fig.~\ref{fig_n02}b,c).
%The maximum distortion of $\langle y(t)\rangle\approx -0.25$ appears at around the same time,
%$t\approx 10$ ns, when $P_2(t)$ reaches its maximum value.
This phase error in the $x$-$y$ plane
arises mainly from the interaction with the second excited state instead of the qubit dissipation~\cite{Lucero10}.

The DRAG method has been theoretically proposed to remove the influence of higher excited states~\cite{MotzoiPRL09,GambettaPRA11}.
As described in Methods, an extra pulse $\bm B_\rmd(t)$ is supplemented to the STA pulse $\bm B(t)$, leading to
the modified field as $\bm B^\pr(t)=\bm B(t)+\bm B_\rmd(t)$.
Under a specified rotating frame characterized by its reference time propagator $\calD(t)$,
the transformed Hamiltonian is factorized as
$H_\calD(t)=\left[\varepsilon(t)I_2+(\hbar/2)\bm B(t)\cdot \bm \sigma\right]\oplus\varepsilon_2(t) |2\rangle\langle 2|$,
where $I_2=|0\rangle\langle0|+|1\rangle\langle1|$ is an identity operator,
and $\varepsilon(t)$ and $\varepsilon_2(t)$ are two shifted energies~\cite{MotzoiPRL09,GambettaPRA11}.
The qubit subspace of $\{|0\rangle, |1\rangle\}$  becomes isolated with the second excited state $|2\rangle$.
The quantum operations of the two-level qubit are accomplished under the $\calD$-frame.
The forms of $\bm B_\rmd(t)$ and $\calD(t)$ are difficult to be solved exactly.
Here we take an approximate DRAG correction~\cite{GambettaPRA11}.  %on the first order of $\Delta^{-1}_2$
By assuming a zero correction along $z$-direction, the DRAG field is analytically written as
$\bm B_{\rmd}(t)=(B_{\rmd; x}(t), B_{\rmd; y}(t), 0)$ with
$B_{\mathrm{d};x}(t) = (1/4\Delta_2)[2\ddot{\theta}(t)-\Omega^2 \sin2\theta(t)]$
and $B_{\mathrm{d};y}(t) = -(\Omega/\Delta_2)\dot{\theta}(t)\cos\theta(t)$. %, respectively.
The detailed $3\times 3$ matrix form of $\calD(t)$ is provided in Supplementary Information.

Next we perform the QST to measure the experimental trajectory subject to the
external field $\bm B^\pr(t)$ after the DRAG correction.
The density matrix  under the $\calD$-frame, $\rho_\calD(t)=\calD^+(t)\rho(t)\calD(t)$,
is calculated using the counterpart $\rho(t)$ under the $\calR$-frame.
For the $3\times 3$ density matrix $\rho(t)$,
an approximation, $\rho_{n2}(t)=\rho_{2n}(t)= \sqrt{P_n(t)P_2(t)}$ with $n=0, 1$,
is applied %for the off-diagonal elements between $\{|0\rangle, |1\rangle\}$ and $|2\rangle$.
%This overestimation
which is acceptable %since $P_2(t)$ is a small number
due to a large anharmonicity parameter $\Delta_2$ and a small value of $P_2(t)$.
The trajectory, $\bm r_\calD(t)=\Tr\{\rho_\calD(t)\bm\sigma\}$, calculated  under the $\calD$-frame
is depicted in Fig.~\ref{fig_n02}a, while the projections $\langle \zeta_\calD(t)\rangle$
along the three directions ($\zeta=x, y, z$) are plotted in Fig.~\ref{fig_n02}b-d respectively.
Compared to the result without the DRAG correction, the phase error in the $x$-$y$ plane
is significantly suppressed. The $x$-projection $\langle x_\calD(t)\rangle$ agrees very
well with the ideal result and the maximum error of
$\langle y_\calD(t)\rangle$  becomes %is largely decreased from 0.25 to
less than 0.05.
The final populations at the two excited states are further improved to $P_1(T_a)=0.95$ and $P_2(T_a)=0.004$.
A fast adiabatic trajectory is thus reliably achieved in our phase qubit with the assistance
of the STA protocol and the DRAG correction. %The fidelity of this fast `adiabatic' state
\\

\noindent {\bf Simulating the topological transition by real-time fast adiabatic trajectories.}
For the SSH model, %in Fig.~\ref{fig_n00},
each unit cell consists of two inequivalent sites
($A$ and $B$, see Supplementary Material). The single-spinless-electron Hamiltonian reads
\be
H=(\hbar\Omega_1/2)\sum_{n} \left( |n, A\rangle\langle n, B|+h.c.\right)+(\hbar\Omega_2/2)\sum_{n}\left( |n, B\rangle\langle n+1, A|+h.c.\right),
\label{eq_n03}
\ee
where $|n, A\rangle$ ($|n, B\rangle$) is the electronic wavefunction of site $A$ ($B$) in the $n$-th
unit cell, and $\Omega_1$ ($\Omega_2$) is twice the intracell (intercell) hopping amplitude in the unit of angular frequency~\cite{Asboth2016short}.
With a periodic boundary condition, the bulk Hamiltonian in Eq.~(\ref{eq_n03}) is block diagonalized
in a quasi-momentum space, i.e., $H=\sum_\theta H(\theta)$. The block element at each quasi-momentum $\theta$
is given by $H(\theta)=(\hbar/2)\bm B_0(\theta)\cdot\bm\sigma$ with $\bm B_0(\theta)= (\Omega_1+\Omega_2\cos \theta, \Omega_2\sin \theta, 0)$ and $0\le \theta \le 2\pi$.
To be consistent with the above adiabatic state transfer, the external field
is rotated to be $\bm B_0(t)=(\Omega_2\sin\theta(t), 0, \Omega_1+\Omega_2\cos\theta(t))$.
The time evolution of $\theta(t)$ mimics a pathway of the quasi-momentum traversing the FBZ.

\begin{figure}[t!]
\includegraphics[width=0.75\linewidth]{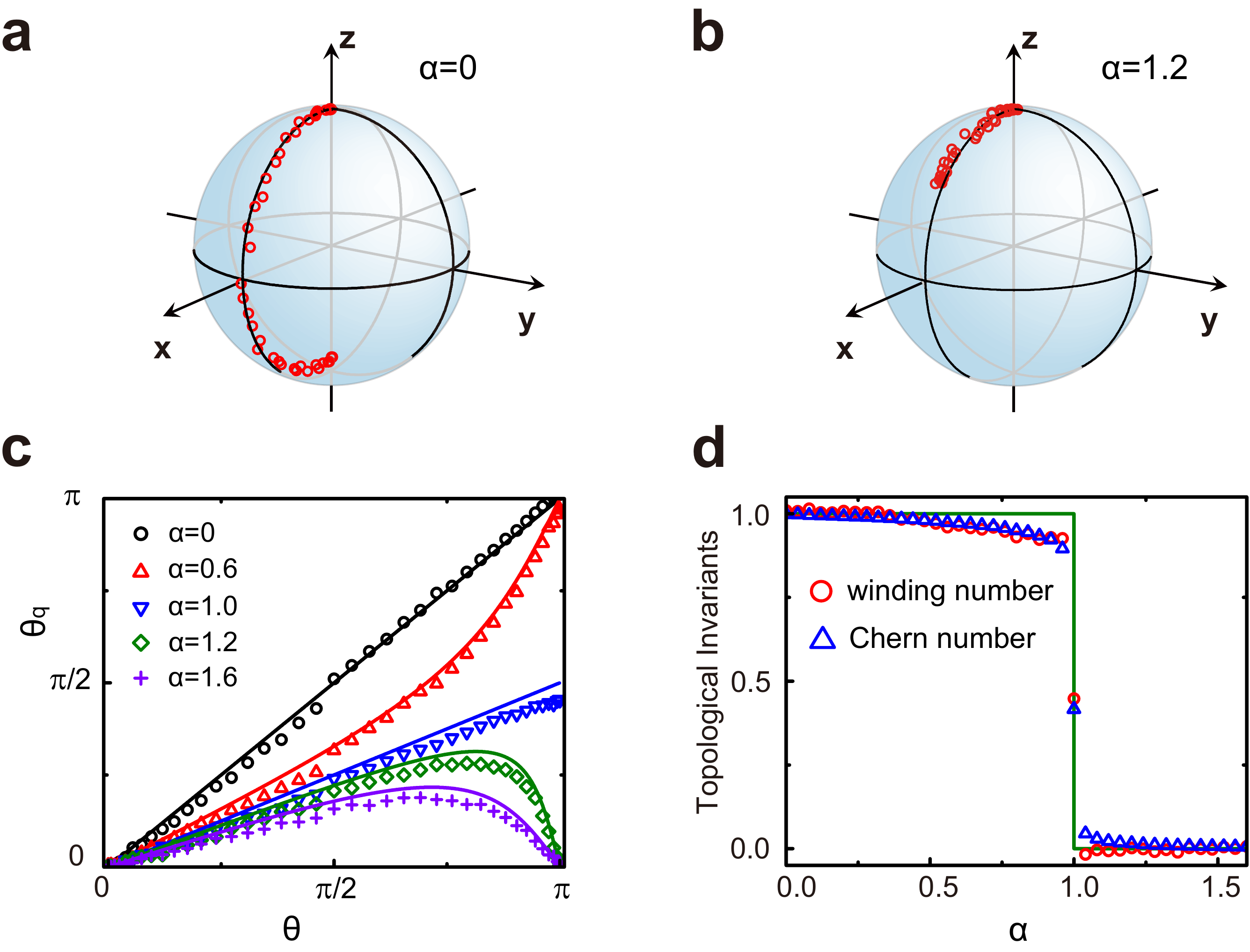}
\caption{
The trajectories of the qubit vector $\bm r_\calD(\theta)|_{\theta=\theta(t)}$ on the Bloch sphere. ({\bf a}) With the hopping amplitude ratio $\alpha=\Omega_1/\Omega_2=0$.  ({\bf b}) With $\alpha=1.2$.
({\bf c}) The polar angle $\theta_{\rm q}$ of the Bloch eigenstate versus the quasi-momentum $\theta$.
From top to bottom, symbols are experimental results of $\alpha=0$ (black circles), $\alpha=0.6$ (red up-triangles),
$\alpha=1$ (blue down-triangles), $\alpha=1.2$ (green diamonds) and $\alpha=1.6$ (purple pluses).
The solid lines are the theoretical predictions under the ideal conditions.
({\bf d}) The topological invariants as the functions of $\alpha$. The red circles and blue up-triangles denote the
experimental result of the winding number for the 1D SSH model and the Chern number for a toy 2D model, respectively.
The green solid line is the theoretical prediction of the ideal topological transitions
in these two systems.}
\label{fig_n03}
\end{figure}

In our experiment, the intercell hopping amplitude is fixed at $\Omega_2/2\pi=30$ MHz while the intracell hopping
amplitude $\Omega_1$ is varied to simulate the topological transition. %from the topologically nontrivial to trivial phases.
The phase qubit is initially reset at the ground state. %, which is the instantaneous spin-up state at $t=0$.
The Hanning-window form, $\theta(t)=(\pi/2)[1-\cos (\pi t/T_a)]$, is chosen for the time evolution of the quasi-momentum.
Due to the intrinsic symmetry of the Hamiltonian, the operation is limited to a half-circle transition with $0\le \theta(t)\le \pi$.
The STA protocol with
the counter-diabatic field $\bm B_\mathrm{cd}(t)$ is applied for a fast adiabatic manipulation, in which
the operation time is set as $T_a=20$ ns.
The DRAG field $\bm B_\mathrm{d}(t)$ is also included to suppress the influence of highly excited states.
The QST measurement is performed every 0.5 ns so that the total $M=41$ quasi-momenta are probed.
The measured density matrix is subsequently transformed into
$\rho_\calD(t)=\calD^+(t)\rho(t)\calD(t)$ under the $\calD$-frame.
%The detailed forms of $\bm B_\mathrm{cd}(t)$, $\bm B_\mathrm{d}(t)$, and $\calD(t)$ are provided in Supporting Information~\cite{xxx}.

For the SSH model, the dispersion relation of the two bulk Bloch bands is unchanged
if the values of $\Omega_1$ and $\Omega_2$ are swapped. However, the topology of the SSH model is
sensitive to the ratio of the two hopping amplitudes, $\alpha=\Omega_1/\Omega_2$.
In the case of $\alpha>1$,  the SSH model behaves as a conventional insulator.
In the opposite case of $\alpha<1$, two eigenstates with almost-zero eigenenergies appear
at the two ends of the SSH model (see Supplementary Information).
Following the bulk-boundary correspondence, the emergence of the edge states can be understood
alternatively from the change of the bulk Bloch eigenstates.
As an illustration, we present the experimental evolutions at two typical ratios, $\alpha=0$ and $\alpha=1.2$.
The trajectories of the qubit vector, $\bm r_\calD(t)=\Tr\{\rho_\calD(t)\bm\sigma\}$, are depicted
in Fig.~\ref{fig_n03}a,b respectively.
As the quasi-momentum evolves from 0 to $\pi$,
the qubit vector evolves from the north to south pole %of the Bloch sphere
for $\alpha=0$ while the qubit vector is retracted to the initial north pole for $\alpha=1.2$.
The separation of the two topological phases are visualized by the different trajectories of $\bm r_\calD(t)$.

The polar angle,
$\theta_{\rm q}(\theta)=\arccos\left[\langle z_\calD(t)\rangle/\sqrt{\langle x_\calD(t)\rangle^2+\langle z_\calD(t)\rangle^2}\right]$,
is experimentally determined to characterize the bulk Bloch eigenstate $|u(\theta)\rangle$. %at the quasi-momentum $\theta(t)$.
The $y$-projection $\langle y_\calD(t)\rangle$ is discarded to reduce the phase error in the estimation. %of $\theta_{\rm q}(\theta)$.
In Fig.~\ref{fig_n03}c, we present the results of $\theta_{\rm q}(\theta)$ for the five hopping amplitude ratios.
Under each condition, the experimental measurement agrees quantitatively well with the theoretical prediction,
$\theta_{\rm q}(\theta)=\arccos\left[(\alpha+\cos \theta)/\sqrt{1+\alpha^2+2\alpha\cos \theta} \right]$.
For $\alpha=0$ and $\alpha=0.6$, the two polar angles %$\theta_{\rm q}(\theta)$
monotonically increase with $\theta$
and reach the almost same value, $\theta_{\rm q}(\theta=\pi)\approx\pi$, at the end of the trajectory.
For $\alpha=1.2$ and $\alpha=1.6$, the two polar angles decrease to zero after an initial increase.
The linear line, $\theta_{\rm q}(\theta)\approx\theta/2$ at $\alpha=1$, represents the transition behavior separating
the two topological phases.
For the other half of the FBZ ($\pi\le \theta\le 2\pi$), the dependence of $\theta_{\rm q}(\theta)$ on the quasi-momentum $\theta$ can be %straightforwardly
deduced using a symmetry argument. The periodic condition of the bulk Bloch eigenstate, $|u(\theta+2\pi)\rangle=|u(\theta)\rangle$,
requires $\theta_{\rm q}(\theta+2\pi)=\theta_{\rm q}(\theta)+2\nu\pi$ with $\nu$ an integer. In the SSH model, this integer is given by
 $\nu=0$ for $\alpha>1$ and $\nu=1$  for $\alpha<1$.
The topological invariant $\nu$ is equivalent to the winding number
of the curve $\bm r_\calD(\theta)$ circulating around the center
of the Bloch sphere. Under the ideal condition, the winding number is defined as
\be
\nu = \frac{1}{2\pi} \int_0^{2\pi} \bm e_y\cdot\left[\tilde{\bm B}_0(\theta)\times d_\theta \tilde{\bm B}_0(\theta)\right] d\theta,
\label{eq_n04}
\ee
where $\tilde{\bm B}_0(\theta)=\bm B_0(\theta)/|\bm B_0(\theta)|$ is a normalized vector and $\bm e_y$ is the
unit vector along the $y$-direction~\cite{Asboth2016short}.
Experimentally, this number is estimated using
$\nu=(1/\pi)[\theta_{\rm q}(\theta=\pi)-\theta_{\rm q}(\theta=0)]$. As shown in Fig.~\ref{fig_n03}d,
a sharp topological transition of the SSH model is identified by our experimental measurement of $\nu$, which is very close
to the theoretical prediction.

Similar to two earlier studies~\cite{RoushanNat14,SchroerPRL14}, the external field can be extended to be
$\bm B_0(\theta, \phi)=(\Omega_2\sin\theta\cos\phi, \Omega_2\sin\theta\sin\phi, \Omega_1+\Omega_2\cos\theta)$,
where the vector of $(\theta, \phi)$ defines a 2D quasi-momentum.
To cover the entire 2D FBZ of the spherical surface,
we can perform the adiabatic state transfer driven by $\bm B_0(\theta(t), \phi(t)=\phi)$
and then the constant azimuthal angle $\phi$ is increased by small steps to form a closed circle of $0\le \phi\le 2\pi$. The Chern number,
\be
\mathcal{C}h = \frac{1}{4\pi}\int\int \tilde{\bm B}_0(\theta, \phi)\cdot \left[\partial_\theta \tilde{\bm B}_0(\theta, \phi)\times \partial_\phi \tilde{\bm B}_0(\theta, \phi)\right] d\theta d\phi,
\label{eq_n05}
\ee
is the topological invariant of 2D systems~\cite{HasanRMP10}. For our given FBZ,
Eq.~\ref{eq_n05} is simplified to be $\mathcal Ch=(1/2)\int_0^\pi \sin\theta_{\rm q}(\theta) (\partial_\theta \theta_{\rm q}) d\theta$,
which can thus be simulated by the adiabatic trajectory subject to the 1D control parameter $\theta(t)$.
In our experiment, we use the data in Fig.~\ref{fig_n03}c to estimate
the Chern number using a summation, $\mathcal Ch=(1/2)\sum_{m=0}^{M-1} [\theta_{\rm q}(m+1)-\theta_{\rm q}(m)]\sin[(\theta_{\rm q}(m+1)+\theta_{\rm q}(m))/2]$.
Figure~\ref{fig_n03}d demonstrates a quantitative agreement between the experimental measurement and the theoretical prediction:
$\mathcal Ch=1$ for $\alpha<1$ and $\mathcal Ch=0$ for $\alpha>1$.
The fast adiabatic trajectories obtained in our 1D experiment thus simulate the topological transition in a simple 2D model.
\\

\begin{figure}[t!]
%\begin{center}
\includegraphics[width=0.75\linewidth]{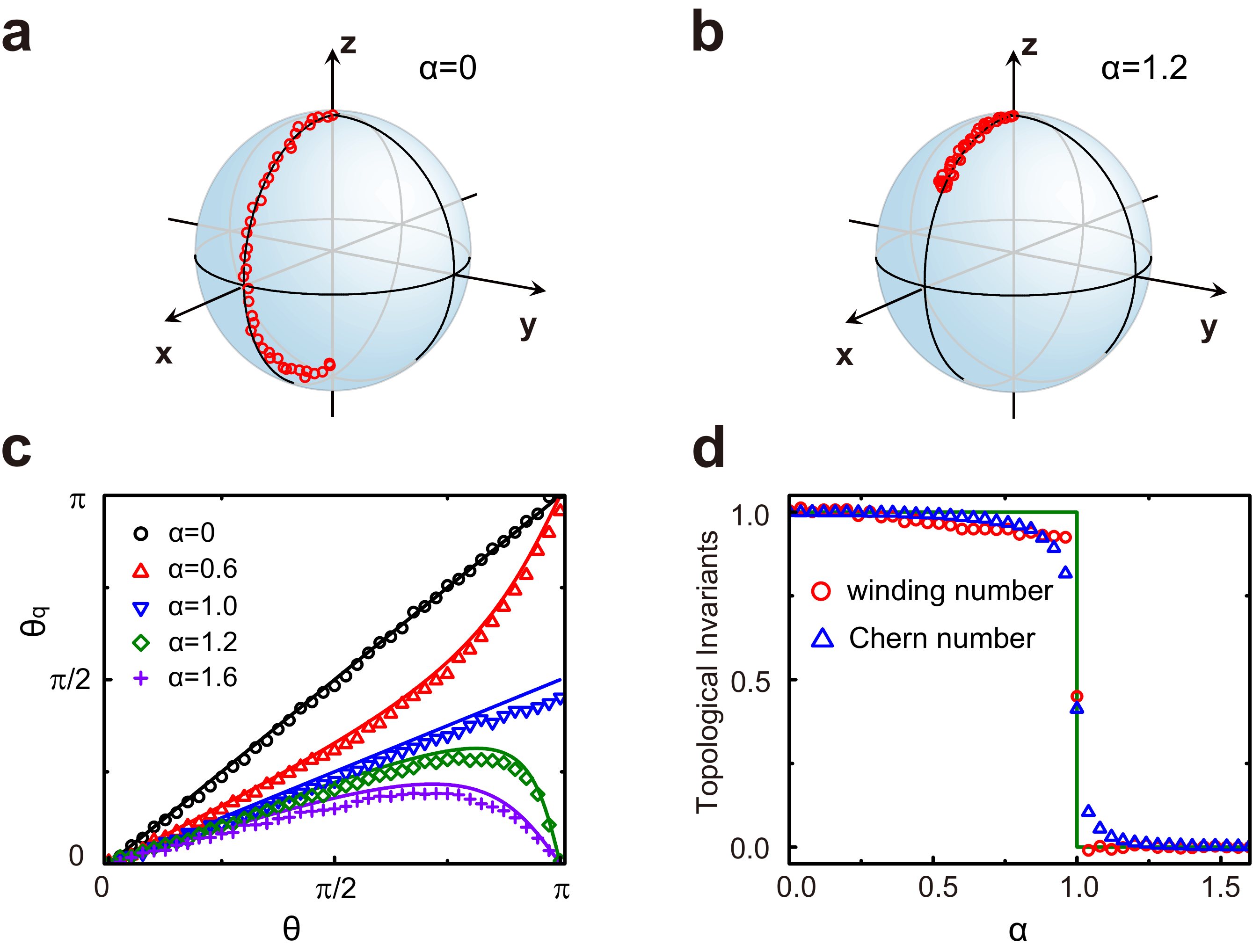}
\caption{
The trajectories of the qubit vector $\bm r_\calD(\theta)|_{\theta=\theta_m(t=T_a)}$ on the Bloch sphere. ({\bf a}) With $\alpha=0$.  ({\bf b}) With $\alpha=1.2$.
({\bf c}) The polar angle $\theta_{\rm q}$ of the bulk Bloch eigenstate versus the quasi-momentum $\theta$.
The parameters are the same as those in Fig.~\ref{fig_n03}c.
({\bf d}) The topological invariants as the functions of $\alpha$.
In {\bf a}-{\bf d}, all the symbols and lines are defined in the same way as their counterparts in Fig.~\ref{fig_n03}.
}
\label{fig_n04}
%\end{center}
\end{figure}

\noindent {\bf Simulating the topological transition by virtual fast adiabatic trajectories.}
For an ideal adiabatic process, each segment can  be regarded as an independent
adiabatic process. A complete adiabatic trajectory is alternatively achieved by
a series of adiabatic state transfers when the control parameter is terminated at intermediate positions
along its designed pathway. Following this simulation scheme, the time evolution of the control parameter is changed to be
$\theta_m(t)=(\pi/2m)[1-\cos(\pi t/T_a)]$, which mimics an even distribution of the quasi-momentum
by $\theta_m(t=T_a)=\pi/m$.
The total $M=41$ ($0\le m\le M$) external fields, $\bm B_0(t)=(\Omega_2\sin\theta_m(t), 0, \Omega_1+\Omega_2\cos\theta_m(t))$,
are applied to generate a virtual adiabatic trajectory in our experiment.
For each $m$-th field, both counter-diabatic and DRAG fields, $\bm B_{\mathrm{cd}}(t)$
and $\bm B_{\mathrm d}(t)$, are supplemented for a fast adiabatic state transfer.
The QST measurement is performed only at the end of the STA operation with $T_a=20$ ns.
In Fig.~\ref{fig_n04}, we present in detail the experimental results based on the virtual trajectories of $\bm r_\calD(\theta)|_{\theta=\theta_m(t=T_a)}$.
All the figures are drawn in the same way as their counterparts in Fig.~\ref{fig_n03} based on
the real-time trajectories of $\bm r_\calD(\theta)|_{\theta=\theta(t)}$. By comparing the results in Fig.~\ref{fig_n03}
and Fig.~\ref{fig_n04}, we find that the accuracies of these two simulation methodologies are close
to each other. Therefore, the virtual adiabatic trajectories
provide an alternative method to reliably simulating the topological transition.
 \\
%The Bloch eigenstates are determined by the virtual adiabatic
%trajectories, and the transitions of the winding number and the Chern number are
%clearly identified as well.

%On one hand, the transition of the Chern number
%is less sharp in the virtual trajectory method, possibly due to the accumulated
%errors from different pulse sequences. On the other hand, the virtual trajectory
%allows a more flexibility of probing quasi-momenta. Since we are interested
%in the final density matrix for each state transfer, the transformation from
%the $\calR$- to $\calD$-frame may not be necessary if $\calD(T_a)=1$ is approximately held.\\

\noindent{\bf Discussion and conclusion} \\
The quantum simulation in this article is built on the technique of fast adiabatic state transfers.
The underlined STA protocol and its extensions have been shown to maintain adiabaticity in a fast operation,
which can help establish practical applications of adiabatic procedures.
As our experiment is focused on a single qubit, it will be worth exploring in the future the STA protocol in
multi-qubit systems.  %~\cite{RoushanNat14}.
The multi-level effect is common in many practical quantum systems.
The DRAG correction applied in our experiment is an efficient way to exclude the influence of
highly excited states. %Compared to earlier studies on the final state, we demonstrated
%a significant improvement on recovering the entire trajectory.
The implementation of the STA protocol and the DRAG correction often requires a complicated drive pulse, which
can however be reliably generated in a superconducting qubit system with a sophisticated
microwave control technique.

In perivous studies of simulating the topological transition, the experiments were performed
by measuring local non-adiabatic responses, integrated phases and transport quantities.
%, which circumvents the difficulty of adiabaticity.
Alternatively, the fast adiabatic evolution of the spin-up state
in our experiment is a simple but transparent method of visualizing the bulk Bloch eigenstates which include all the
geometric and topological information.
The sharp transitions in our measurements of the winding number
and the Chern number verify a strong protection of topological phases unless the energy gap is closed and reopened
around the transition point. The virtual construction of the adiabatic trajectory
in addition to the real-time approach provides the flexibility in mimicking the FBZ.
Although a 1D quasi-momentum is considered in our study, this simulation method can be straightforwardly
extended to a realistic 2D system. After collecting multiple adiabatic trajectories, we may obtain
the Bloch eigenstates over the entire 2D FBZ, which will be explored in the future.
\\

\noindent{\bf Methods} \\
\noindent{\bf Experimental setup.}
The superconducting phase qubit  used in this experiment is the same as that
in our previous experiment~\cite{ZZXPRA17}. An anharmonic LC resonator
is formed by  an Al/AlOx/Al Josephson junction coupled with a  parallel capacitor ($C_\mathrm{q} = 1$ pF) and a loop inductor ($L_\mathrm{q} = 720$ pH).
The lowest two energy levels ($\varepsilon_0$ and $\varepsilon_1$) are used as the ground and excited states of a qubit,
and their resonance frequency is  $\omega_{10}=(\varepsilon_1-\varepsilon_0)/\hbar$.
The second excited state with energy $\varepsilon_2$ induces an ahhamonic frequency shift,
$\Delta_2=\omega_{21}-\omega_{10} = (\varepsilon_2+\varepsilon_0-2\varepsilon_1)/\hbar$.
In our experiment, these two parameters are given by $\omega_{10}/{2\pi}=5.7$ GHz and $\Delta_2/{2\pi}=-200$ MHz.
The microwave drive signal is synthesized by an IQ mixer in which
two low-frequency quadratures are mixed with a local oscillator signal.
A on-chip superconducting quantum interference device (SQUID) is used to measure the population of various qubit states
(see Supplementary Information). The density matrix is further extracted by the QST method. \\

\noindent{\bf Counter-diabatic field in the STA protocol.}
A non-degenerate reference Hamiltonian $H_0(t)$ is diagonalized in the instantaneous eigen basis set,
leading to $H_0(t)=\sum_n \varepsilon_n(t) |n(t)\rangle\langle n(t)|$ where $|n(t)\rangle$ is the $n$-th instantaneous eigenstate and $\varepsilon_n(t)$ is its eigenenergy.
In a slow adiabatic operation, a quantum system initially at $|n(t=0)\rangle$ can remain
at this eigenstate. The system wavefunction is given by $|\psi(t)\rangle = \exp[i\varphi(t)]|n(t)\rangle$,
where $\varphi(t)$ is an accumulated phase. The non-adiabatic transitions can however destroy
adiabaticity. To speed-up the adiabatic operation while maintaining adiabaticity, the STA protocol
requires an additional counter-diabatic Hamiltonian $H_{\mathrm{cd}}(t)$.
The wavefunction is expanded as $|\psi(t)\rangle = \sum_n a_n(t) |n(t)\rangle$.
With respect to the total Hamiltonian, $H(t)=H_0(t)+H_{\mathrm{cd}}(t)$, the time evolution of each coefficient $a_n(t)$ follows
\be
\hbar\dot{a}_n(t) &=& -i \left[ \varepsilon_n(t)-i\hbar\langle n(t)|\partial_t n(t)\rangle \right]a_n(t) - i  \langle n(t) |H_{\mathrm{cd}}(t)|n(t)\rangle a_n(t) \no \\
& & -i\sum_{m(\not=n)} \left[-i\hbar\langle n(t)|\partial_t m(t)\rangle +\langle n(t)|H_{\mathrm{cd}}(t)|m(t)\rangle\right] a_m(t).
\label{eq_M01}
\ee
To recover the adiabatic evolution under the reference Hamiltonian, the constraints,
\be
\langle n(t) |H_{\mathrm{cd}}(t)|m(t)\rangle =\left\{\ba{ll} 0  &~~\mathrm{for}~~~m= n  \\
 i\hbar\langle n(t)|\partial_t m(t)\rangle &~~~\mathrm{for}~~~m\neq n \ea  \right.,
\label{eq_M02}
\ee
are required for the counter-diabatic Hamiltonian, which is satisfied by
\be
H_{\mathrm{cd}}(t) = i \hbar\sum_n \big[ |\partial_t n(t)\rangle\langle n(t)|- \langle n(t)|\partial_t n(t)\rangle |n(t)\rangle \langle n(t)|\big].
\label{eq_M03}
\ee
For a spin-half particle, the reference Hamiltonian in general follows
$H_0(t)=\hbar \bm B_0(t)\cdot\bm \sigma/2$
with $\bm B_0(t)=|\bm B_0(t)|(\sin\theta(t)\cos\phi(t), \sin\theta(t)\sin\phi(t), \cos\theta(t))$.
The amplitude of this external field, $|\bm B_0(t)|$, gives rise to the two eigenenergies,
$\varepsilon_{\uparrow, \downarrow}(t) =  \pm\hbar |\bm B_0(t)|/2$, %and $\varepsilon_{\downarrow}(t) = - \hbar |\bm B_0(t)|/2$,
while the two angle variables, $\{\theta(t), \phi(t)\}$, determine the two eigenstates,
\be
\left\{\ba{ccc}
|s_{\uparrow}(t)\rangle &=& \cos\left[\frac{\theta(t)}{2}\right] |0\rangle + \sin\left[\frac{\theta(t)}{2}\right]e^{i\phi(t)} |1\rangle  \\
|s_{\downarrow}(t)\rangle &=& -\sin\left[\frac{\theta(t)}{2}\right]e^{-i\phi(t)} |0\rangle + \cos\left[\frac{\theta(t)}{2}\right] |1\rangle  \ea \right. .
\label{eq_M032}
\ee
Next we substitute Eq.~\ref{eq_M032} into Eq.~\ref{eq_M03} and obtain the three elements of the counter-diabatic field, which are given by
\be
\left\{ \ba{ccl} B_{\mathrm{cd}; x}(t) &=&   -\dot{\theta}(t)\sin\phi(t)-\dot{\phi}(t)\sin\theta(t)\cos\theta(t)\cos\phi(t) \\
B_{\mathrm{cd}; y}(t) &=&  \dot{\theta}(t)\cos\phi(t)-\dot{\phi}(t)\sin\theta(t)\cos\theta(t)\sin\phi(t) \\
B_{\mathrm{cd}; z}(t) &=&  \dot{\phi}(t)\sin^2\theta(t) \ea\right. .
\label{eq_M033}
\ee
Equation~\ref{eq_M033} can be further rewritten in a cross product form as
\be
\bm B_\mathrm{cd}(t) = \frac{1}{|\bm B_0(t)|^2}\bm B_0(t)\times\dot{\bm B}_0(t),
\label{eq_M04}
\ee
which is always orthogonal to the reference field $\bm B_0(t)$.
For the special case of $\bm B_0(t)=(\Omega\sin\theta(t), 0, \Omega\cos\theta(t))$, the counter-diabatic field is written as $\bm B_{\mathrm{cd}}(t)= (0, \dot{\theta}(t), 0)$.
\\

\noindent{\bf Population evolution of the second excited state.}   The anharmonicity in our phase qubit cannot be
ignored and the framework of a three-level system is necessary. Under the $\calR$-frame, the three-level Hamiltonian of an anharmonic oscillator
is given by $H(t)=(\hbar/2)\bm B(t)\cdot \bm S+\hbar\Delta_2|2\rangle\langle 2|$
with the STA field $\bm B(t)=\bm B_0(t)+\bm B_{\mathrm{cd}}(t)=(\Omega\sin\theta(t), \dot{\theta}(t), \Omega\cos\theta(t))$.
A diagonalization is applied to the qubit subspace, $\{|0\rangle, |1\rangle\}$, giving the instantaneous spin-up
($|s_\uparrow(t)\rangle$) and spin-down ($|s_\downarrow(t)\rangle$) states subject to the reference field $\bm B_0(t)$. %in the framework of the two-level system.
The time-varying basis set is changed to $\{|s_\uparrow(t)\rangle, |s_\downarrow(t)\rangle, |2\rangle\}$,
and the Hamiltonian is transformed accordingly. If the qubit is designed to follow the adiabatic
trajectory of the spin-up state, we can focus on the subspace of $\{|s_\uparrow(t)\rangle, |2\rangle\}$
and omit $|s_\downarrow(t)\rangle$. The partial Hamiltonian governing the state evolution within this subspace
is given by
\be
H(t) = (\hbar \Omega/2)|s_\uparrow(t)\rangle\langle s_\uparrow(t)|+\hbar\Delta^\prime_2(t) |2\rangle\langle 2|
+ \hbar\left[ J_{\uparrow 2}(t) |s_\uparrow(t)\rangle\langle 2| + h.c.\right],
%\left( \ba{cc}  \frac{\Omega}{2}  & \frac{\sqrt{2}\sin[\theta(t)/2]\left[\Omega\sin\theta(t)-i\dot{\theta}(t)\right]}{2} \\
%\frac{\sqrt{2}\sin[\theta(t)/2]\left[\Omega\sin\theta(t)+i\dot{\theta}(t)\right]}{2} &  -\Delta_2-\frac{3\Omega\cos\theta(t)}{2}\ea \right).
\label{eq_M05}
\ee
with $\Delta^\prime_2(t) = \Delta_2-(3/2)\Omega\cos\theta(t)$ and $J_{\uparrow 2}(t)=(\sqrt{2}/2)\sin[\theta(t)/2]\left[\Omega\sin\theta(t)-i\dot{\theta}(t)\right]$.
Next we assume an adiabatic evolution beginning with the initial state $|s_\uparrow(t=0)\rangle=|0\rangle$.
The population of the second excited state is then approximated as
\be
P_2(t) &\approx& \frac{1}{2} \left[1-\frac{|\Delta^\prime_2(t)-\Omega/2|}{\sqrt{[\Delta^\prime_2(t)-\Omega/2]^2+4|J_{\uparrow 2}(t)|^2}} \right] \no \\
&\approx& \frac{ |J_{\uparrow 2}(t)|^2}{[\Delta^\prime_2(t)-\Omega/2]^2},
\label{eq_M06}
\ee
where the second equation is obtained under the consideration of $|\Delta_2|\gg \Omega, |\dot{\theta}(t)|$. For a polar angle with $\theta(t=T_a)=\pi$,
the final population of the second excited state becomes $P_2(t=T_a)\approx\dot{\theta}(t=T_a)^2/[2(\Delta_2+\Omega)^2]$  . \\

\noindent{\bf First-order approximation of the DRAG correction.} For a multi-level anharmonic oscillator,
the DRAG method is proposed to isolate the qubit subspace $\{|0\rangle, |1\rangle\}$ from
higher excited states (detailed in Supplementary Information). In the case of the three-level system,
the above STA Hamiltonian is expanded over a perturbation parameter $\Delta^{-1}_2$,  giving $H(t)=H^{(-1)}(t)+H^{(0)}(t)$
with $H^{(-1)}(t)=\hbar\Delta_2|2\rangle\langle 2|$ and $H^{(0)}(t)=(\hbar/2)\bm B(t)\cdot \bm S$. The DRAG field $\bm B_{\rm d}(t)$
is responsible for higher order corrections, $H_{\rm d}(t)=H^{(1)}(t)+\cdots=
(\hbar/2)[\bm B^{(1)}_{\rm d}(t)+\cdots]\cdot \bm S$.
The modified total Hamiltonian is written as $H^\prime(t)=H(t)+H_{\rm d}(t)$.
%after the DRAG correction.
The time evolution of this three-level system is inspected under an alternative $\calD$-frame,
which is defined by its reference time propagator, $\calD(t)=\exp[-i \calM(t)]$.
The same expansion over $\Delta^{-1}_2$ is applied to the exponent term $\calM(t)$, giving
$\calM(t)=\calM^{(1)}(t)+\calM^{(2)}(t)+\cdots$. The transformation from the original
$\calR$-frame to the new $\calD$-frame changes  the total Hamiltonian to be
\be
H_\calD(t)=\calD^+(t)H^\prime(t)\calD(t)+i\dot{\calD}^+(t)\calD(t),
\label{eq_M07}
\ee
and the density matrix follows $\rho_\calD(t)=\calD^+(t)\rho(t)\calD(t)$.

Under the construction of the DRAG method, we expect a factorized form,
$H_\calD(t)=[\varepsilon(t)I_2+(\hbar/2)\bm B(t)\cdot \bm \sigma]\oplus\varepsilon_2(t) |2\rangle\langle 2|$,
for the transformed Hamiltonian. The two energies also follow the expansion over $\Delta^{-1}_2$,
giving $\varepsilon(t)=\varepsilon^{(1)}(t)+\cdots$ and
 $\varepsilon_2(t)=\hbar\Delta_2+\varepsilon^{(0)}_2(t)+\varepsilon^{(1)}_2(t)+\cdots$.
Two additional constraints, $\calD(t=0)=1$ and $\calD(t=T_a)=1$,
are considered so that the $\calD$-frame recovers the $\calR$-frame at the initial
and final moments of the operation. Next we expand both sides of
Eq.~\ref{eq_M07} over $\Delta^{-1}_2$ order by order, which results in
a series of equations for $\{\bm B^{(i)}_{\rm d}(t)\}$ and
$\{\calM^{(i)}(t)\}$.
These equations are however difficult to be solved exactly.
In our experiment, we truncate the expansion of Eq.~\ref{eq_M07} up to the first order of $\Delta^{-1}_2$.
%together with an assumption of a zero correction along the $z$-direction ($B^{(1)}_z(t)=0$).
Consequently, we obtain $\bm B_{\rm d}(t)\approx\bm B^{(1)}_{\rm d}(t)$
and $\calD(t)\approx\exp\{-i[\calM^{(1)}(t)+\calM^{(2)}(t)]\}$ while the explicit forms of
the three perturbations are provided in Supplementary Information. \\

\noindent{\bf Acknowledgements}\\
The work reported here is supported by the National Basic Research Program of China (2014CB921203, 2015CB921004),
the National Natural Science Foundation of China (NSFC-11374260, NSFC-21173185),
and the Fundamental Research Funds for the Central Universities in China (2016XZZX002-01).
Devices were made at John Martinis's group using equipments of UC Santa Barbara Nanofabrication Facility,
a part of the NSF-funded National Nanotechnology Infrastructure Network. \\

\noindent{\bf Conflicts of interest} \\
The authors declare that they have no competing interests.

\noindent{\bf References} \\

\end{document}

\title{Supplementary Material for ``Simulating a Topological Transition in a Superconducting Phase Qubit by Fast Adiabatic Trajectories''}
\author{Tenghui Wang}
\affiliation{Physics Department, Zhejiang University, Hangzhou, 310027, China}
\author{Zhenxing Zhang}
\affiliation{Physics Department, Zhejiang University, Hangzhou, 310027, China}
\author{Liang Xiang}
\affiliation{Physics Department, Zhejiang University, Hangzhou, 310027, China}
%\author{{\color{red}(Jiadong Yao)}}
%\affiliation{Physics Department, Zhejiang University, Hangzhou, 310027, China}
\author{Zhihao Gong}
\affiliation{Physics Department, Zhejiang University, Hangzhou, 310027, China}
\author{Jianlan Wu}
\affiliation{Physics Department, Zhejiang University, Hangzhou, 310027, China}
\author{Yi Yin \footnote{Correspondence and requests for materials should be addressed to Y.Y. (email: yiyin@zju.edu.cn) or to J.L.W.
(email: jianlanwu@zju.edu.cn).}}
\affiliation{Physics Department, Zhejiang University, Hangzhou, 310027, China}
\affiliation{Collaborative Innovation Center of Advanced Microstructures, Nanjing, 210093, China}

\maketitle

\section{Readout of a Superconducting Phase Qubit}
\label{sec:s1}

\subsection{Population Measurment}
\label{sec:s1a}

\begin{figure}
\centering
\includegraphics[width=0.55\columnwidth]{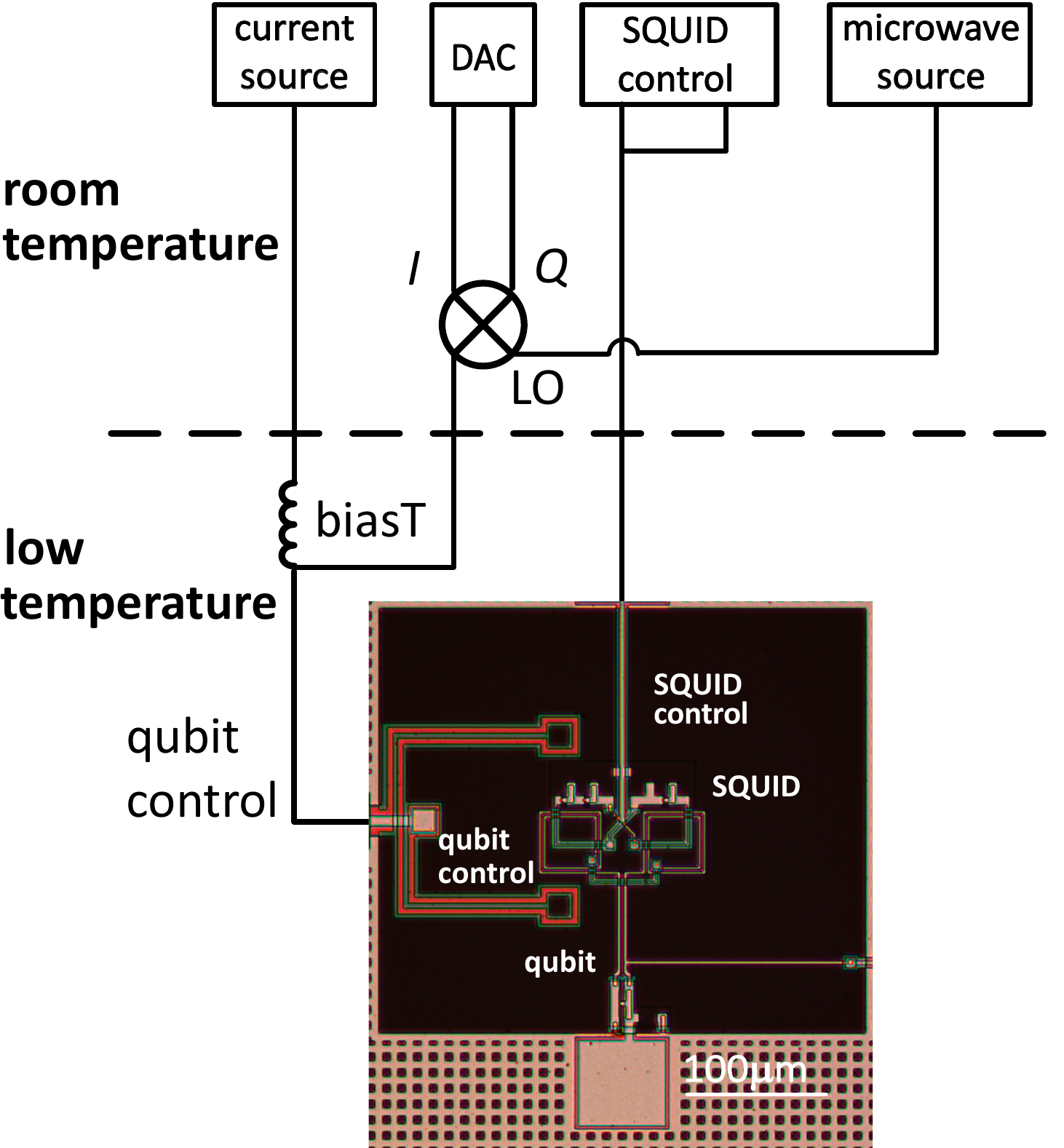}
\caption{A schematic diagram of our experimental setup of a superconducting phase qubit: external electronic controls at room temperature
         and a chip sample placed in a dilution refrigerator with a base temperature $\sim 10$ mK. A photomicrograph of the chip sample
         is provided to show a qubit, a SQUID, and their control lines connected to the room-temperature electronics.
}
\label{fig:s1}
\end{figure}

\begin{figure}
\centering
\includegraphics[width=0.7\columnwidth]{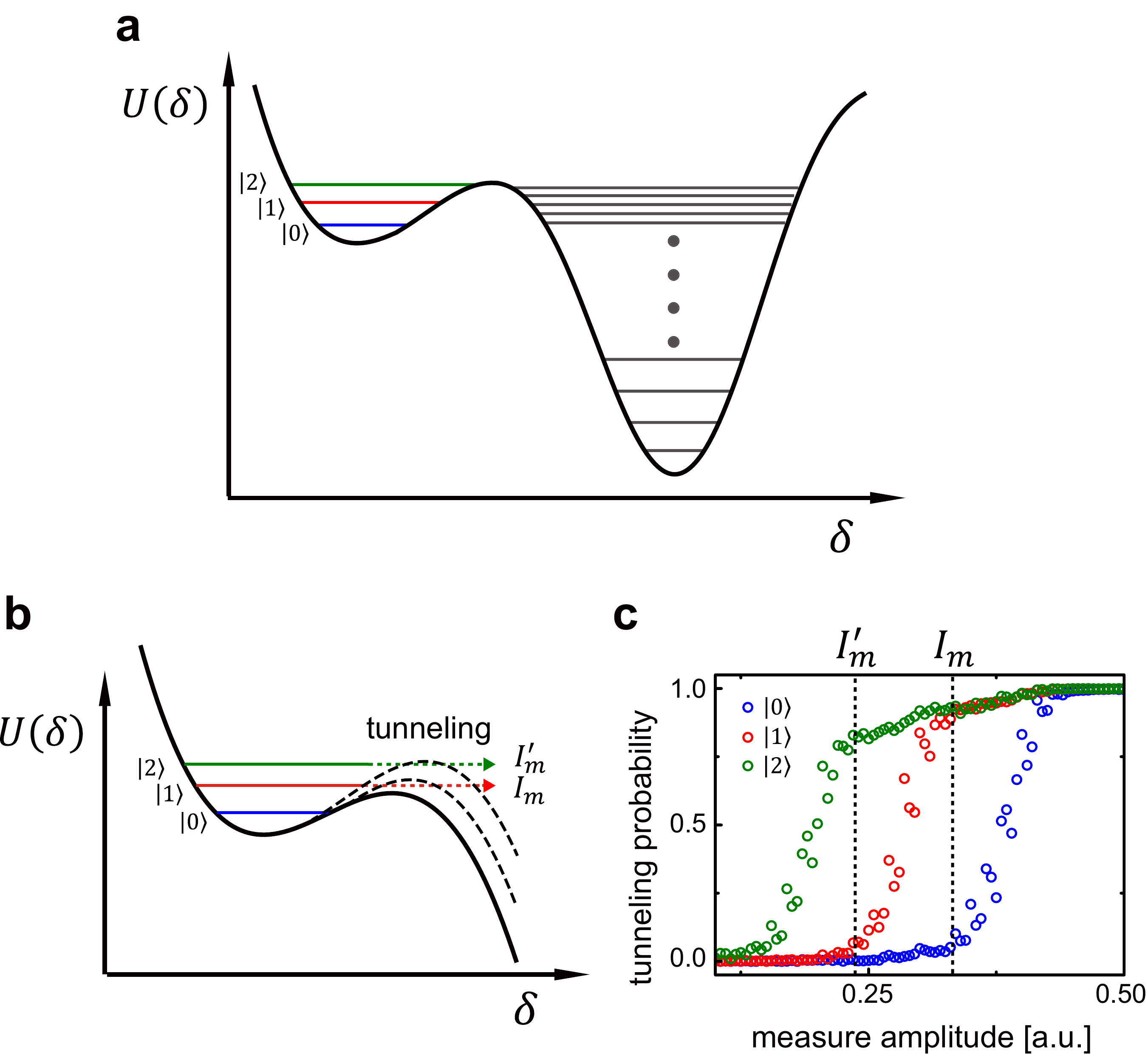}
\caption{Readout of a superconducting phase qubit. (a) The energy potential $U(\delta)$ as a function of the phase difference $\delta$ across the Josephson junction. %The quantum manipulation
         %is applied to energy levels in the left shallow well.
         (b) The populations of various excited states are selected to tunnel to the right well under different
         measurement current bias $I_{\rm m}$. (c) The tunneling probability as a function of the measurement current bias for the three lowest  energy levels, $\{|0\rangle, |1\rangle, |2\rangle\}$.
         In our experiment, the two specific current biases, $I_{\rm m}$ and $I^\pr_{\rm m}$, are chosen for measuring the quantum state (see text for details).}
\label{fig:s2}
\end{figure}

In Fig.~\ref{fig:s1}, we present a schematic diagram of our experimental setup
and a photomicrograph of our chip sample~\cite{MartinisReview,ZZXPRA17}.
The main components of the chip are a superconducting phase qubit,
a superconducting quantum interference device (SQUID) and their control lines.
The phase qubit circuit is composed of a Josephson junction
($I_0 =2$ $\mu$A), a loop inductance ($L_\mathrm{q} = 720$ pH) and
a capacitance ($C_\mathrm{q} = 1$ pF).
Figure~\ref{fig:s2}a shows the energy potential of this qubit loop
as a function of the  phase difference across the Josephson junction.
An anharmonic resonator is formed in the left well,  where
the lowest two energy levels are chosen as the ground ($|0\rangle$) and excited ($|1\rangle$) states
of a phase qubit. Since the quantized energy levels ($\varepsilon_n$)  %in the left well
are not equally spaced,
a quantum manipulation can be efficiently restricted within the subspace of $\{|0\rangle, |1\rangle\}$.
However, the interaction with higher excited states ($|n\ge 2\rangle$) induces
population leakage and a phase error in the qubit subspace.
In our experiment, the three lowest energy levels
are considered. The transition frequency between the ground
and first excited states is $\omega_{10}/{2\pi}=5.7$ GHz
and the transition frequency between the first and second excited
states is $\omega_{21}/{2\pi}= 5.5$ GHz.

A key step of controllable quantum manipulation is to read out the quantum state
of the qubit. In our phase qubit system, the quantum state in the left well
can tunnel through the energy barrier and relax in the right well (Fig.~\ref{fig:s2}b),
inducing a different flux which is detectable using the coupled SQUID.
 A pulse of the current bias $I$ is applied to lower the potential barrier and increase
the tunneling probability. After repeating the same measurement $\sim 10^3$ times,
we obtain the tunneling probability $P_{\rm t}$ of this quantum state.
By preparing the quantum state at each $n$-th energy level, we
draw a calibration curve, $f_n(I)=P_{\rm t}(I)|_{|\Psi\rangle=|n\rangle}$, as shown in Fig.~\ref{fig:s2}c.
For a general quantum state, $|\Psi\rangle=\sum_{n} c_n |n\rangle$,
 the population $P_n=|c_n|^2$ at each $n$-th state is calculated from
these calibration curves. Here we consider the two-level
qubit as an example. A specific current bias $I_{\rm m}$ is selected from which
the tunneling probabilities of $|0\rangle$ and $|1\rangle$ are most separated, giving $f_0=6.5\%$
and $f_1=92.5\%$ respectively. The tunneling probability $P_{\rm t}$ measured at $I_{\rm m}$
is an average result from the two quantum states, i.e.,
$P_{\rm t} =  P_0 f_0 + P_1 f_1$, where $P_0$ and $P_1$ are the actual
populations at the ground and excited states respectively.
Together with the normalization condition, $P_0+P_1=1$,
the two populations are solved by
\be
\left(\ba{c} P_0\\ P_1 \ea \right)=
\left(\ba{cc} f_0&f_1\\1&1 \ea\right)^{-1} \left(\ba{c} P_{\rm t}\\1 \ea\right).
\label{eq_S1}
\ee
The same measurement method can be extended to multi-level systems.
For the three-level system (qutrit),
the two measure current biases, $I_{\rm m}$ and $I^\pr_{\rm m}$,
are applied to distinguish the three energy levels (Fig.~\ref{fig:s2}b).
From the two extracted tunneling probabilities, $P_{\rm t}$ and $P^\pr_{\rm t}$,
the actual populations at the three quantum states are calculated as
\be
\left(\ba{c} P_0\\ P_1 \\ P_2 \ea \right)=
\left(\ba{ccc} f_0&f_1 & f_2 \\  f^\pr_0 & f^\pr_1 & f^\pr_2 \\ 1 & 1 & 1 \ea\right)^{-1} \left(\ba{c} P_{\rm t}\\ P^\pr_{\rm t} \\1 \ea\right),
\label{eq_S2}
\ee
where the two sets of the calibration tunneling
probabilities are $\{f_0=6.5\%, f_1=92.5\%, f_2=93\%\}$
and $\{f'_0=0.3\%, f'_1=9.3\%, f'_2=83.1\%\}$ in our experiment.

\subsection{Quantum State Tomography (QST) Measurement}
\label{sec:s1b}

To fully determine a quantum state, we need to extract the information of coherence in addition to population.
For the quantum state of a two-level qubit, the density matrix (either a pure or mixed state) is expanded as
\be
\rho = \frac{1}{2}\left(I_2 + x \sigma_x + y \sigma_y + z \sigma_z\right),
\label{eq_S03}
\ee
with $I_2=|0\rangle\langle 0|+|1\rangle\langle 1|$. Here we introduce three Pauli operators,
$\sigma_x = |0\rangle\langle1|+|1\rangle\langle 0|$, $\sigma_y = -i|0\rangle\langle1|+i|1\rangle\langle 0|$,
and $\sigma_z = |0\rangle\langle0|-|1\rangle\langle 1|$. The three projections, $x$, $y$ and $z$
along the three directions, determine a vector, $\bm r = (x, y, z)$, which is named as the Bloch vector. Together with a vector
of Pauli operators, $\bm \sigma=(\sigma_x, \sigma_y, \sigma_z)$, the density matrix in Eq.~(\ref{eq_S03})
is rewritten as $\rho = (I_2+\bm r\cdot \bm \sigma)/2$. The Bloch vector of a pure state
stays on the surface of the Bloch sphere ($|\bm r|=1$),
while that of a mixed state is inside the Bloch sphere ($|\bm r|<1$).
%The density matrix is determined if the three elements of the Bloch vector are known.
The $z$-projection, $z=P_0-P_1$, is extracted from
the population measurement in Section~\ref{sec:s1a}. To extract the $x$-projection, we rotate
the quantum state by an angle of $-\pi/2$ around the $y$-axis and the density matrix is changed to be
\be
\rho^\pr = U_y(-\pi/2) \rho U^+_y(-\pi/2),
\label{eq_S04}
\ee
where $U_\zeta(\theta)=\exp[-i\theta\sigma_\zeta/2]$ is an unitary operator for a rotation angle of $\theta$
around the $\zeta(=x, y, z)$-axis. Experimentally, the $U_\zeta(\theta)$ gate is realized by a Gaussian-envelop
pulse with the frequency $\omega_{10}$.
The rotated density matrix in Eq.~(\ref{eq_S04}) is given by $\rho^\pr=(I_2-z\sigma_x+y\sigma_y+x\sigma_z)/2$. The
population measurement determines the $x$-projection, $x=P^\pr_0-P^\pr_1$, where $P^\pr_0$ and $P^\pr_1$
are the populations of the ground and excited states in the rotated density matrix.
In practice, the $U_y(\pi/2)$ gate is also applied for the measurement
and the $x$-projection is averaged from the results under the operations of the
$U_y(\pi/2)$ and $U_y(-\pi/2)$ gates. The $y$-projection is similarly obtained using the operations
of the $U_x(\pi/2)$ and $U_x(-\pi/2)$ gates.

The above QST method can be extended to multi-level systems, where the density
matrix is written as
\be
\rho = \sum_n P_n |n\rangle\langle n| + \sum_{m<n} \frac{1}{2}\left(x_{mn} \sigma_{mn; x}+y_{mn} \sigma_{mn; y}\right).
\label{eq_S05}
\ee
Here we introduce two operators, $\sigma_{mn; x}=|m\rangle\langle n|+|n\rangle\langle m|$
and $\sigma_{mn; y}=-i|m\rangle\langle n|+i|n\rangle\langle m|$, for an arbitrary pair of quantum states, $\{|m\rangle, |n\rangle\}$.
The off-diagonal elements, $\rho_{mn}=\left(x_{mn}- i y_{mn}\right)/2$  with $m<n$, can be similarly extracted by rotating the quantum state
under unitary operators, $U_{mn; \zeta}(\theta)=\exp[-i \theta\sigma_{mn; \zeta}/2]$ with $\zeta=x, y$.
For the three-level system, the operation of $U_{01; y}(-\pi/2)$ on the density matrix leads to
\be
\rho^\pr = U_{01; y}(-\pi/2) \rho U^+_{01; y}(-\pi/2),
\label{eq_S06}
\ee
where the three transformed population elements are
\be
\left\{ \ba{lll} P^\pr_0 &=& \frac{1}{2}(P_0+P_1+x_{01}) \\
P^\pr_1 &=& \frac{1}{2}(P_0+P_1-x_{01}) \\
P^\pr_2 &=& P_2
\ea \right. .
\label{eq_S07}
\ee
The measurement method in Section~\ref{sec:s1a} can be used to determine $x_{01}$.
The other off-diagonal elements, $y_{01}$, $x_{12}$ and $y_{12}$, are similarly extracted.
In principle, this method can be applied to the coherence between the ground and second excited ($|2\rangle$)
states, which is however limited
by the range of the microwave drive frequency.
%{\color{red}(by the phase qubit system)}.
Instead, we can apply two operator products, $V_{02; x}=U_{12; x}(\pi/2)U_{01; x}(\pi)$ and $V_{02; y}=U_{12; y}(\pi/2)U_{01; x}(\pi)$.
The population elements under the unitary transformation of $V_{02; x}$ are
\be
\left\{ \ba{lll} P^\pr_0 &=& P_1  \\
P^\pr_1 &=& \frac{1}{2}(P_0+P_2+x_{02}) \\
P^\pr_2 &=& \frac{1}{2}(P_0+P_2-x_{02})
\ea \right. ,
\label{eq_S08}
\ee
while the population elements under the unitary transformation of $V_{02; y}$ are
\be
\left\{ \ba{lll} P^\pr_0 &=& P_1  \\
P^\pr_1 &=& \frac{1}{2}(P_0+P_2+y_{02}) \\
P^\pr_2 &=& \frac{1}{2}(P_0+P_2-y_{02})
\ea \right. .
\label{eq_S09}
\ee
Then, $x_{02}$ and $y_{02}$ are extracted from the population measurement method.
In our experiment, the population of the second excited state is in general a small number,
$P_2\lesssim 0.05$, so that the coherence elements, $|\rho_{02}|$ and $|\rho_{12}|$, are
also relatively small. In an approximate but acceptable manner, we only perform
the QST to extract $\rho_{01}$ and $\rho_{10}$. The other
coherence elements are estimated from population elements,
i.e., $\rho_{02}=\rho_{20} \approx\sqrt{P_0 P_2}$
and $\rho_{12}=\rho_{21}\approx\sqrt{P_1 P_2}$. %The Lindbald simulation shows that
The error induced by this approximation is small when we
focus on the $x$- and $z$-elements of the Bloch vector after the derivative removal by adiabatic gates (DRAG)
correction.

\section{Hamiltonian in the Rotating Frame}
\label{sec:s2}

For a $(N+1)$-level anharmonic oscillator driven by an external pulse $\lambda(t)$, its Hamiltonian under the laboratory frame ($\calL$-frame or the Schr\"{o}dinger picture)
is written as
\be
H_{\calL}(t) = \sum_{n=0}^N\varepsilon_n |n\ra\la n| + \hbar\lambda(t) \sum_{n=0}^{N-1}\sqrt{n+1}(|n\ra\la n+1|+|n+1\rangle\langle n|),
\label{eq_S10}
\ee
where $\varepsilon_n$ is energy of the $n$-th quantum state. In comparison with a harmonic oscillator,
we define an anharmonicity parameter, $\Delta_n = (\varepsilon_n-\varepsilon_0)/\hbar-n\omega_{10}$, for each $n(\ge 2)$-th excited state.
%where $\omega_{10}=(\varepsilon_1-\varepsilon_0)/\hbar$  is the transition frequency between the ground ($|0\ra$) and first excited ($|1\ra$) states.
The classical microwave pulse $\lambda(t)$ follows a general form as
\be
\lambda(t) &=& E(t) \cos[\omega_d t+\Phi(t)],
\label{eq_S11}
\ee
where $E(t)$, $\omega_d$ and $\Phi(t)$ are the amplitude, frequency and phase factor of this drive pulse.

Next we build a rotating frame ($\calR$-frame) based on a reference time propagator,
\be
\calR(t)= \exp\left\{-i\sum_{n=0}^N\left(n-\frac{1}{2}\right)\left[\omega_d t+\xi(t)\right]|n\ra\la n|  \right\},
\label{eq_S12}
\ee
where $\xi(t)$ is a phase factor to create a time-dependent detuning. Another
phase factor, $\phi(t)=\xi(t)-\Phi(t)$, is subsequently introduced.
Following the construction of the interaction picture, the Hamiltonian under the $\calR$-frame is written as
\be
H_{\calR}(t) = \calR^+(t)H_{\calL}(t)\calR(t)+i \dot{\calR}^+(t)\calR(t),
\label{eq_S13}
\ee
where the second term on the right hand side (RHS) arises from the time evolution of the $\calR$-frame.
By substituting Eqs.~(\ref{eq_S10}) and~(\ref{eq_S12}) into Eq.~(\ref{eq_S13}), we obtain the transformed Hamiltonian as
\be
H_{\calR}(t) &=& \left(\varepsilon_0+\hbar\frac{\omega_{10}}{2}\right)I_{N+1}+\sum_{n=0}^N \left(n-\frac{1}{2}\right)\hbar\left[\omega_{10}-\omega_d-\dot{\xi}(t)\right]|n\ra\la n| \no \\
&&+\sum_{n=2}^N\hbar\Delta_{n} |n\ra\la n|+\frac{\hbar}{2}\sum_{n=0}^{N-1} E(t)e^{-i\left[\xi(t)-\Phi(t)\right]}\sqrt{n+1}|n\ra\la n+1| \no \\
&&+\frac{\hbar}{2}\sum_{n=0}^{N-1} E(t)e^{i\left[\xi(t)-\Phi(t)\right]}\sqrt{n+1}|n+1\ra\la n|,
\label{eq_S14}
\ee
where $I_{N+1}=\sum_{n=0}^N |n\rangle\langle n|$ is an identity operator %for the $(N+1)$-level anharmonic oscillator
and the rotating wave approximation (RWA) is applied for the last two terms on the RHS.
In our experiment, we set the drive frequency of the
external pulse to be the same as the transition frequency of the qubit, i.e., $\omega_d=\omega_{10}$.
As a result, Eq.~(\ref{eq_S14}) is simplified to be
\be
H_{\calR}(t)
&=& \left(\varepsilon_0+\hbar\frac{\omega_{10}}{2}\right)I_{N+1} +\frac{\hbar}{2}\bm B(t)\cdot \bm S +\sum_{n=2}^N\hbar\Delta_{n} |n\ra\la n|,
\label{eq_S16}
\ee
where $\bm B(t)=(E(t)\cos\phi(t), E(t)\sin\phi(t), \dot{\xi}(t))$ is  an effective magnetic field and $\bm S= (S_x, S_y, S_z)$ is a
vector of three ladder operators,
\be
\left\{\ba{ccl}
S_x &=& \sum_{n=0}^{N-1}\sqrt{n+1}\left(|n+1\rangle\langle n|+|n\rangle\langle n+1|\right) \\
S_y &=& \sum_{n=0}^{N-1}\sqrt{n+1}\left(i|n+1\rangle\langle n|-i|n\rangle\langle n+1|\right)  \\
S_z &=& \sum_{n=0}^N (1-2n) |n\rangle\langle n| \ea \right.
\label{eq_S17}
\ee
%with $h.c.$ denoting the Hermitian conjugate.
The first term on the RHS of Eq.~(\ref{eq_S16}) is often ignored since
this constant energy shift does not affect the time evolution of the system. In our experiment, the sophisticated microwave
technique allows an accurate generation of a designed external field, $\bm B(t)=(B_x(t), B_y(t), B_z(t))$,
by choosing appropriate functions of $E(t)$, $\phi(t)$ and $\xi(t)$. In addition to the Hamiltonian,
the density matrix under the $\calR$-frame, $\rho_{\calR}(t)=\calR^+(t)\rho_{\calL}(t)\calR(t)$,
 is transformed from its counterpart $\rho_{\calL}(t)$ under the $\calL$-frame.
If higher excited states can be efficiently suppressed
and the system recovers the two-level qubit, Eq.~(\ref{eq_S16}) is further reduced to
\be
H_{\mathcal R}(t) = \frac{\h}{2} \bm{B}(t)\cdot\bm{\sigma}.
\label{eq_S18}
\ee
This Hamiltonian describes
a spin-half particle driven by an external magnetic field $\bm B(t)$,
which can be used for the quantum simulation of a non-interacting two-band system.

In the main text, we start directly from the transformed $\calR$-frame instead of the $\calL$-frame,
since the $\calR$-frame provides a more transparent physical picture of our
experiments. Accordingly, we will omit the index $\calR$ in the Hamiltonian
and the density matrix, i.e., $H_{\calR}(t)\rightarrow H(t)$ and $\rho_{\calR}(t)\rightarrow \rho(t)$.

\section{Derivative Removal by Adiabatic Gates (DRAG) Correction}
\label{sec:s3}

In this section, we review the DRAG method, which follows closely the derivation in Ref.~\cite{GambettaPRA11}.
The DRAG method is proposed to exclude the influence of higher excited states and recover the two-level qubit from a multi-level
anharmonic oscillator. For simplicity, we only take the three-level anharmonic oscillator as an example.
As demonstrated in Section~\ref{sec:s2}, the three-level Hamiltonian under the $\calR$-frame is written  as
\be
H(t) =  \frac{\h}{2} \bm{B}(t)\cdot\bm{S}+\h\Delta_2|2\ra\la 2|,
\label{eq_S19}
\ee
with $\bm B(t) = (B_x(t), B_y(t), B_z(t))$.
To remove the coupling between the first and second excited states, we introduce an additional field,
$\bm B_{\rm d}(t)=(B_{\rmd; x}(t), B_{\rmd; y}(t), B_{\rmd; z}(t))$,
to modify the Hamiltonian as follows,
\be
H^\pr(t)= \frac{\h}{2}\bm B^\pr(t)\cdot \bm S+\hbar \Delta_2 |2\rangle\langle 2|
\label{eq_S20}
\ee
with $\bm B^\pr(t)=\bm B(t)+\bm B_{\rm d}(t)$.
Besides, we create a new rotating frame ($\calD$-frame) based on a reference time propagator, $\calD(t)=\exp[-i \calM(t)]$,
where the exponent term $\calM(t)$ is a Hermitian operator. The Hamiltonian
under the $\calD$-frame is transformed into
\be
H_\calD(t)=\calD^+(t)H^\pr(t)\calD(t)+i\dot{\calD}^+(t)\calD(t).
\label{eq_S21}
\ee
Following the design of the DRAG method, the transformed Hamiltonian $H_\calD(t)$  is required to be factorized into
\be
H_\calD(t) =\left[\varepsilon(t)I_2+ \frac{\hbar}{2}\bm B(t)\cdot \bm \sigma\right] \oplus \varepsilon_2(t)|2\rangle\langle 2|,
\label{eq_S22}
\ee
where %$I_2=|0\rangle\langle0|+|1\rangle\langle1|$ is an identity operator, and
$\varepsilon(t)$ and $\varepsilon_2(t)$ are two shifted energies. In Eq.~(\ref{eq_S22}),
the first part on the RHS is implemented in the qubit subspace of $\{|0\rangle, |1\rangle\}$, while the second part %, $H_{\calD; II}(t)=\varepsilon_2(t)|2\rangle\langle 2|$,
in the subspace of $\{|2\rangle\}$. Therefore, we can effectively recover the two-level qubit if the initial
state is prepared in the subspace of $\{|0\rangle, |1\rangle\}$. In the original proposal of the DRAG method, the time propagator
is further required to satisfy
\be
\calD(t=0)=1~~~\mathrm{and}~~~\calD(t=T_a)=1,
\label{eq_S23}
\ee
so that the $\calD$-frame recovers the $\calR$-frame
at the initial and final moments of the quantum operation. In addition to the Hamiltonian, the
density matrix $\rho_{\calD}(t)$ under the $\calD$-frame is
%transformed from its counterpart $\rho(t)$ under the $\calR$-frame,
given by
\be
\rho_{\cal D}(t) = \calD^+(t)\rho(t)\calD(t).
\label{eq_S24}
\ee

Although  the constrains of Eqs.~(\ref{eq_S22}) and~(\ref{eq_S23}) can be
realized by more than one choices of $\bm B_{\rmd}(t)$ and $\calM(t)$,
the DRAG correction is difficult to be solved exactly.
Here we follow the approach in Ref.~\cite{GambettaPRA11} by assuming a large anharmoncity, i.e., $\Delta_2 \gg |\bm B(t)|$.
The DRAG field $\bm B_{\rm d}(t)$ and the exponent term $\calM(t)$ are then decomposed into
\be
\bm B_{\rm d}(t) &=& \bm B^{(1)}_{\rm d}(t)+\bm B^{(2)}_{\rm d}(t)+\cdots , \label{eq_S24a}\\
\calM(t)         &=& \calM^{(1)}(t)+\calM^{(2)}(t)+\cdots,
\label{eq_S25}
\ee
where $\bm B^{(j)}(t)$ and $\calM^{(j)}(t)$ on the $j$-th expansion order
are proportional to $(1/\Delta_2)^j$.
The Hamiltonian under the $\calR$-frame is expanded as
\be
H^\prime(t) &=& H^{(-1)}(t)+H^{(0)}(t)+H^{(1)}(t)+\cdots \no \\
&=& \hbar\Delta_2|2\rangle\langle2|  + (\hbar/2) \bm B(t)\cdot \bm S + (\hbar/2) \bm B^{(1)}_{\rm d}(t) \cdot \bm S +\cdots,
\label{eq_S26}
\ee
while the $\calD(t)$-operator is expanded as
\be
\calD(t) &=& \exp\left\{-i \left[ \calM^{(1)}(t)+\calM^{(2)}(t)+\cdots \right] \right\}  \no \\
&=& 1-i\calM^{(1)}(t)-\left[\frac{\left(\calM^{(1)}(t)\right)^2}{2}+i \calM^{(2)}(t)  \right]+\cdots.
\label{eq_S27}
\ee
For simplicity, all the diagonal elements are assumed to be zero in $\calM^{(j)}(t)$, i.e.,
\be
\calM^{(j)}(t) = \sum_{m< n} \calM^{(j)}_{mn; x}(t) (|m\rangle\langle n|+|n\rangle\langle m|)+\calM^{(j)}_{mn; y}(t)(-i|m\rangle\langle n|+i|n\rangle\langle m|).
%~~~\mathrm{for}~~~j=1, 2, \cdots.
\label{eq_S28}
\ee
Next we substitute Eqs.~(\ref{eq_S26}) and~(\ref{eq_S27}) into Eq.~(\ref{eq_S21}), which gives rise to
\be
H_{\calD}(t) &=& H^{(-1)}(t)+ \left\{H^{(0)}(t)+i\left[\calM^{(1)}(t), H^{(-1)}(t)\right]  \right\} \no \\
& & +\left\{H^{(1)}(t)-\frac{1}{2}\left[\calM^{(1)}(t),\left[\calM^{(1)}(t), H^{(-1)}(t)\right]\right] + i\left[\calM^{(2)}(t), H^{(-1)}(t)\right]\right. \no \\
& & ~~~\left. +i\left[\calM^{(1)}(t), H^{(0)}(t)\right]-\dot{\calM}^{(1)}(t) \right\}+\cdots.
\label{eq_S29}
\ee
By matching Eqs.~(\ref{eq_S29}) and~(\ref{eq_S22}) order by order, we obtain a series of equations,
\be
{\rm minus~first~order}: && \hbar\Delta_2|2\rangle\langle2|,  \label{eq_S30} \\
{\rm zeroth~order}: && (\hbar/2)\bm B(t)\cdot \bm \sigma +\varepsilon^{(0)}_2 |2\rangle\langle 2| \no \\
                   &=& (\hbar/2)\bm B(t)\cdot \bm S+i\left[\calM^{(1)}(t), \hbar\Delta_2|2\rangle\langle2|\right],  \label{eq_S31} \\
{\rm first~order}: && \varepsilon^{(1)}(t)\left(|0\rangle\langle 0|+|1\rangle\langle 1|\right)+\varepsilon^{(1)}_2 |2\rangle\langle 2| \no \\
                   &=& (\hbar/2) \bm B^{(1)}_{\rm d}(t) \cdot \bm S -\frac{1}{2}\left[\calM^{(1)}(t),\left[\calM^{(1)}(t), H^{(-1)}(t)\right]\right] \no \\
                   & +& i\left[\calM^{(2)}(t), \hbar\Delta_2|2\rangle\langle2|\right]+i\left[\calM^{(1)}(t), (\hbar/2) \bm B(t)\cdot \bm S\right]-\dot{\calM}^{(1)}(t), \label{eq_S32} \\
& \vdots & \no
\ee
For the zeroth order Hamiltonian $H^{(0)}_{\calD}(t)$, Eq.~(\ref{eq_S31}) is reached under the following conditions,
\be
\left\{\ba{llllll}
\calM^{(1)}_{02; x}(t)&=& 0~~~~&\calM^{(1)}_{02; y}(t) &=& 0 \\
\calM^{(1)}_{12; x}(t)&=&\frac{1}{\sqrt{2}\Delta_2}B_y(t) ~~~
&\calM^{(1)}_{12; y}(t)&=&-\frac{1}{\sqrt{2}\Delta_2}B_x(t)\\
\varepsilon^{(0)}_2(t) &=& - \frac{3\hbar}{2} B_z(t)   & & &  \ea \right. .
\label{eq_S33}
\ee
For the first order Hamiltonian $H^{(1)}_{\calD}(t)$, the total 9 equations can be derived from Eq.~(\ref{eq_S32}),
while the total 11 variables need to be determined. To solve this discrepancy, we introduce additional constrains.
The condition, $[H^{(1)}_{\calD}(t)]_{0, 0}=[H^{(1)}_{\calD}(t)]_{1, 1}$,  in Eq.~(\ref{eq_S32}) leads to
\be
B^{(1)}_{\rmd; z}(t)
&=&\sqrt{2} B_{y}\left[\sqrt{2}\calM^{(1)}_{01;x}(t)-\frac{1}{2}\calM^{(1)}_{12;x}(t)\right]
-\sqrt{2}B_{x}\left[\sqrt{2}\calM^{(1)}_{01;y}(t)-\frac{1}{2}\calM^{(1)}_{12;y}(t)\right],
\label{eq_S34}
\ee
which is solved by a special set, $\{B^{(1)}_{\rmd; z}(t)=0,
\calM^{(1)}_{01;x}(t)=\calM^{(1)}_{12;x}(t)/2\sqrt{2}, \calM^{(1)}_{01;y}(t)=\calM^{(1)}_{12;y}(t)/2\sqrt{2}\}$.
The remaining 8 variables are then fully determined~\cite{GambettaPRA11}. Here we summarize this specific DRAG correction
for $H_\calD(t)$ up to the first order of $1/\Delta_2$: The first order DRAG field reads
\be
\left\{\ba{ccl} B^{(1)}_{\rmd; x}(t) &=& \frac{1}{2\Delta_2}\left[\dot{B}_{y}(t)-B_{z}(t)B_{x}(t)\right] \\
B^{(1)}_{\rmd; y}(t) &=& -\frac{1}{2\Delta_2}\left[\dot{B}_{x}(t)+B_{z}(t)B_{y}(t)\right] \\
B^{(1)}_{\rmd; z}(t) &=& 0 \ea \right. .
\label{eq_S35}
\ee
The first order exponent term is composed of
\be
\left\{\ba{llllll}
\calM^{(1)}_{01;x}(t)&=&\frac{1}{4\Delta_2}B_y(t)~~~~  &\calM^{(1)}_{01;y}(t)&=&-\frac{1}{4\Delta_2}B_x(t) \\
\calM^{(1)}_{12; x}(t)&=&\frac{1}{\sqrt{2}\Delta_2}B_y(t) &\calM^{(1)}_{12; y}(t)&=&-\frac{1}{\sqrt{2}\Delta_2}B_x(t) \\
\calM^{(1)}_{02; x}(t)&=& 0 &\calM^{(1)}_{02; y}(t) &=& 0  \ea \right. ,
\label{eq_S36}
\ee
while the relevant elements of the second order exponent term are composed of
\be
\left\{\ba{llllll}
\calM^{(2)}_{12; x}(t)&=&-\frac{1}{\sqrt{2}\Delta_2}B^{(1)}_y(t)~~~~&\calM^{(2)}_{12; y}(t)&=&\frac{1}{\sqrt{2}\Delta_2}B^{(1)}_x(t) \\
\calM^{(2)}_{02; x}(t)&=&\frac{3}{4\sqrt{2}\Delta^2_2}B_x(t)B_y(t)~~~~&\calM^{(2)}_{02; y}(t)&=&\frac{3}{8\sqrt{2}\Delta^2_2}\left[B^2_y(t)-B^2_x(t)\right]
\ea \right. .
\label{eq_S37}
\ee
The energy shift of the ground and first excited states is
\be
\varepsilon^{(1)}(t) = -\frac{\hbar}{4\Delta_2}\left[B^2_x(t)+B^2_y(t)\right],
\label{eq_S38}
\ee
while the energy shift of the second excited state is
\be
\varepsilon^{(0)}_2(t) = - \frac{3\hbar}{2} B_z(t),~~~     \varepsilon^{(1)}_2(t) = -\frac{3\hbar}{2} B^{(1)}_{\rmd; z}(t)+\frac{\hbar}{2\Delta_2}\left[B^2_x(t)+B^2_y(t)\right].
\label{eq_S39}
\ee

In our experiment, we apply the DRAG correction of  $\bm B_{\rmd}(t) = \bm B^{(1)}_{\rmd}(t)$ and $\calM(t)=\calM^{(1)}(t)+\calM^{(2)}(t)$.
To satisfy the initial and final conditions in Eq.~(\ref{eq_S23}),
the solutions in Eqs.~(\ref{eq_S35})-(\ref{eq_S39})
imposes additional constrains, $B_x(t=0)=B_x(t=T_a)=0$ and $B_y(t=0)=B_y(t=T_a)=0$.

\section{The Explicit Formulation of the  STA protocol and DRAG correction in our experiment}
\label{sec:s4}

In this section, we summarize the counter-diabatic field, the DRAG field, and the DRAG time propagator
used in our experiment. The reference field is written as
$\bm B_0(t) = (B_{0; x}(t), 0, B_{0; z}(t))$ with $B_{0; x}(t)=\Omega_2\sin\theta(t)$
and $B_{0; z}(t)=\Omega_1+\Omega_2\cos\theta(t)$.
The counter-diabatic field for the STA protocol is calculated using
\be
\bm B_{\rm cd}(t) = \frac{1}{|\bm B_0(t)|^2} \bm B_0(t) \times \dot{\bm B}_0(t),
\label{eq_S40}
\ee
which leads to $\bm B_{\rm cd}(t) = (0, B_{{\rm cd}; y}(t), 0)$.
The nonzero element of the counter-diabatic field is $B_{{\rm cd}; y}(t) = \dot{\theta}_{\rm q}(t)$,
and the angle $\theta_{\rm q}(t)$ is defined as
\be
\theta_{\rm q}(t) = \left\{\ba{lll}\arctan [B_{0; x}(t)/B_{0; z}(t)]~~~ &\mathrm{if}& B_{0; z}(t)\ge 0   \\
 \pi+\arctan [B_{0; x}(t)/B_{0; z}(t)] ~~~ &\mathrm{if}&  B_{0; z}(t)<0 \ea \right.  .
\label{eq_S41}
\ee
In Eq.~(\ref{eq_S41}), we assume two positive hopping amplitudes ($\Omega_1>0$ and $\Omega_2>0$)
and the polar angle $\theta(t)$ in the range of $[0, \pi]$.

Following Eq.~(\ref{eq_S35}), we calculate the first order DRAG field as
$\bm B_{\rmd}(t)=(B_{\rmd; x}(t), B_{\rmd; y}(t), 0)$,
where the two nonzero elements are
\be
\left\{\ba{ccl} B_{\rmd; x}(t) &=& \frac{1}{2\Delta_2}\left\{\ddot{\theta}_{\rm q}(t)-\Omega_2\sin\theta(t)[\Omega_1+\Omega_2\cos\theta(t)]\right\} \\
B_{\rmd; y}(t) &=& -\frac{1}{2\Delta_2}\left\{\Omega_2\cos\theta(t)\dot{\theta}(t)+[\Omega_1+\Omega_2\cos\theta(t)]\dot{\theta}_{\rm q}(t)\right\} \ea \right. .
\label{eq_S42}
\ee
Following Eqs.~(\ref{eq_S36}) and~(\ref{eq_S37}), we obtain nonzero elements of the first and second order exponent terms,
\be
\left\{\ba{llllll} \calM^{(1)}_{01; x}(t) &=& \frac{1}{4\Delta_2}\dot{\theta}_{\rm q}(t)~~~&\calM^{(1)}_{01; y}(t) &=& -\frac{1}{4\Delta_2}\Omega_2\sin\theta(t)  \\
\calM^{(1)}_{12; x}(t) &=& \frac{1}{\sqrt{2}\Delta_2}\dot{\theta}_{\rm q}(t)~~~&\calM^{(1)}_{12; y}(t) &=& -\frac{1}{\sqrt{2}\Delta_2}\Omega_2\sin\theta(t)   \\
\calM^{(2)}_{12; x}(t) &=& -\frac{1}{\sqrt{2}\Delta_2}B_{\rmd; y}(t)~~~&\calM^{(2)}_{12; y}(t) &=& \frac{1}{\sqrt{2}\Delta_2}B_{\rmd; x}(t)  \\
\calM^{(2)}_{02; x}(t) &=& \frac{3}{4\sqrt{2}\Delta^2_2}\Omega_2\sin\theta(t)\dot{\theta}_{\rm q}(t)~~~
&\calM^{(2)}_{02; y}(t) &=& \frac{3}{8\sqrt{2}\Delta^2_2}\left[\dot{\theta}^2_{\rm q}(t)-\Omega^2_2\sin^2\theta(t)\right]
\ea \right.
\label{eq_S43}
\ee
The time propagator is then given by $\calD(t)=\exp[-i \calM(t)]$ with $\calM(t)=\calM^{(1)}(t)+\calM^{(2)}(t)$.
In our experiment, we consider two kinds of time evolutions for the polar angle: a linear increase, $\theta(t)=(\pi/T_a)t$,
and a Hanning-window form, $\theta(t)=(\pi/2)[1-\cos(\pi t/T_a)]$. Only the second one satisfies
the constrains, $B_{0; x}(t=0)=B_{0; x}(t=T_a)=0$ and $B_{{\rm cd}; y}(t=0)=B_{{\rm cd}; y}(t=T_a)=0$,
at the initial and final moments. Thus, the DRAG correction in Eq.~(\ref{eq_S42}) and~(\ref{eq_S43})
is applied to the Hanning-window pulse.

\section{Introduction to Topological Phases}
\label{sec:s5}

\subsection{The Su-Schrieffer-Heeger (SSH) model}
\label{sec:s5a}

\begin{figure}%[t!]
\includegraphics[width=0.65\linewidth]{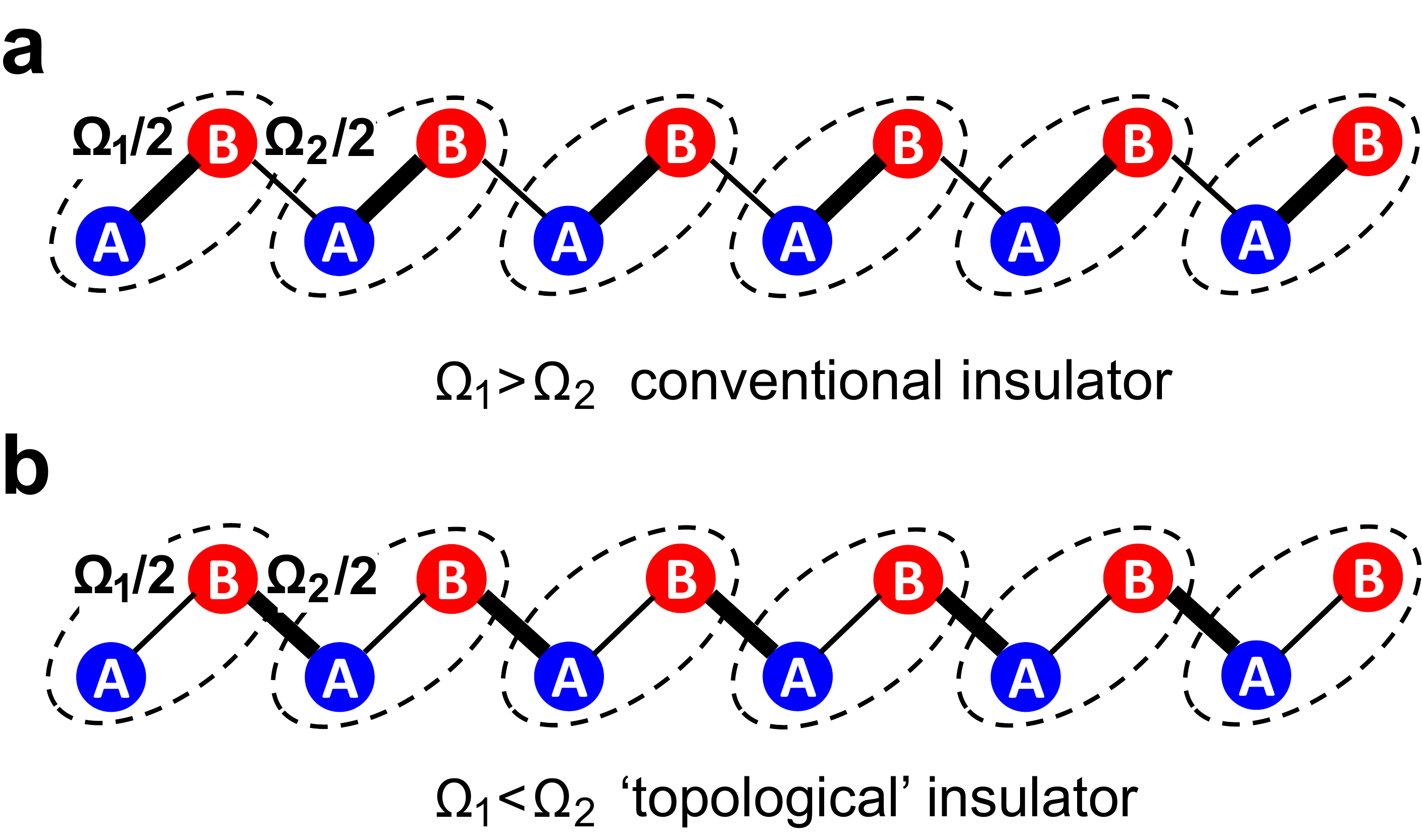}
\caption{{\bf Two topological phases of the SSH model.}
({\bf a}) A conventional insulator from $\Omega_1>\Omega_2$. ({\bf b}) A `topological' insulator from $\Omega_1<\Omega_2$.
Here $\Omega_1/2$ ($\Omega_2/2$) is the intracell (intercell) hopping amplitude.
Each dashed ellipse represents a unit cell consisting
of two sites, $A$ and $B$.
}
\label{fig_n00}
\end{figure}

In this subsection, we briefly review the background of the Su-Schrieffer-Heeger (SSH) model~\cite{SSHmodel79,Asboth2016short}.
The polyacetylene is an insulator instead of a conductor due to the Peierls instability induced by
the interaction between electrons and lattice vibrations. The spontaneous dimerization
of the polyacetylene is described by the one-dimensional (1D)
SSH model with staggered hopping amplitudes.
As shown in Fig.~\ref{fig_n00}, a 1D lattice contains  $N$ unit cells, each consisting of
two inequivalent sites, $A$ and $B$. The Hamiltonian of the SSH model is given by
\be
H = \sum_{n=1}^N \left[\frac{\hbar\Omega_1}{2}|n,A\ra\la n,B|+h.c.\right]+\sum_{n=1}^{N-1} \left[\frac{\hbar\Omega_2}{2}|n+1,A\ra\la n,B|+h.c.\right],
\label{eq_S44}
\ee
where $|n,A\ra$ ($|n, B\rangle$) is the single-electron wave function in site $A$ ($B$) of the $n$-th unit cell
and $\Omega_1$ ($\Omega_2$) is twice the intracell (intercell) hopping amplitude in the unit of angular frequency.
In Eq.~(\ref{eq_S44}), we ignore the possible next nearest neighboring interaction, which is crucial
in the two-dimensional (2D) Haldane model~\cite{Haldane}.

At the first step, a periodic boundary condition is assigned to the SSH model and
the Hamiltonian in  Eq.~(\ref{eq_S44}) is rewritten as
\be
H &=& \sum_{n=1}^N \left[\frac{\hbar\Omega_1}{2}|n,A\ra\la n,B|+h.c.\right] \no \\
&&+\sum_{n=1}^{N} \left[\frac{\hbar\Omega_2}{2}|{\rm mod}(n+1, N),A\ra\la n,B|+h.c.\right].
\label{eq_S45}
\ee
For two sub-lattices formed by sites $A$ and $B$ separately,
we define two Fourier-transformed  wavefunctions,
\be
|\theta_k, A\ra &=& \frac{1}{\sqrt{N}}\sum_{n=1}^Ne^{in\theta_k} |n,A\ra,  \label{eq_S46} \\
|\theta_k, B\ra &=& \frac{1}{\sqrt{N}}\sum_{n=1}^Ne^{in\theta_k} |n,B\ra,  \label{eq_S47}
\ee
where $\theta_k=2k\pi/N$ ($k=0, 1, \cdots, N-1$) is a quasi-momentum in the unit of radian.
In our quantum simulation of real-time trajectories, the quasi-momentum $\theta(t)$ is not equally spaced
in the range of $[0, 2\pi)$, which however does not affect our conclusion since $\theta_k$
becomes continuous as $N\rightarrow \infty$. By substituting
the inverse Fourier transforms, $|n,A\ra=(1/\sqrt{N})\sum_{k=0}^{N-1}\exp(-in\theta_k) |\theta_k, A\rangle$
and $|n, B\ra=(1/\sqrt{N})\sum_{k=0}^{N-1}\exp(-i n\theta_k) |\theta_k, B\rangle$, into Eq.~(\ref{eq_S44}),
the bulk Hamiltonian is rewritten as
\be
H  = \frac{\hbar }{2}\sum_{k=0}^{N-1} \left[(\Omega_1+\Omega_2e^{-i\theta_k}) |\theta_k, A\ra\la \theta_k, B|+h.c.\right],
\label{eq_S48}
\ee
which is block diagonalized with respect to each quasi-momentum, i.e., $H=\sum_{k=0}^{N-1} H(\theta_k)$.
From now on, we omit the index $k$ of the quasi-momentum $\theta_k$ due to the one-to-one mapping between
these two variables. Equation~(\ref{eq_S48})
represents a typical non-interacting two-band Hamiltonian. In the basis set of $\{|\theta, A\rangle, |\theta, B\rangle\}$,
the Hamiltonian $H(\theta)$ is written in a $2\times 2$ matrix form, given by
\be
H(\theta)= \frac{\hbar}{2} \left(\ba{ll}
0 & \Omega_1+\Omega_2e^{-i\theta}\\
\Omega_1+\Omega_2e^{i\theta} & 0 \ea \right).
\label{eq_S49}
\ee
With an effective magnetic field, $\bm B_0(\theta)=(\Omega_1+\Omega_2\cos\theta, \Omega_2\sin\theta, 0)$,
Eq.~(\ref{eq_S49}) is further simplified to be $H(\theta)=\hbar \bm B_0(\theta)\cdot\bm \sigma/2$.
Therefore, the pathway of the quasi-momentum traversing the first Brillouin zone (FBZ, $0\le \theta < 2\pi$)
is equivalently described by a motion of $\theta(t)$ evolving over time.

\begin{figure}
\centering
\includegraphics[width=0.8\columnwidth]{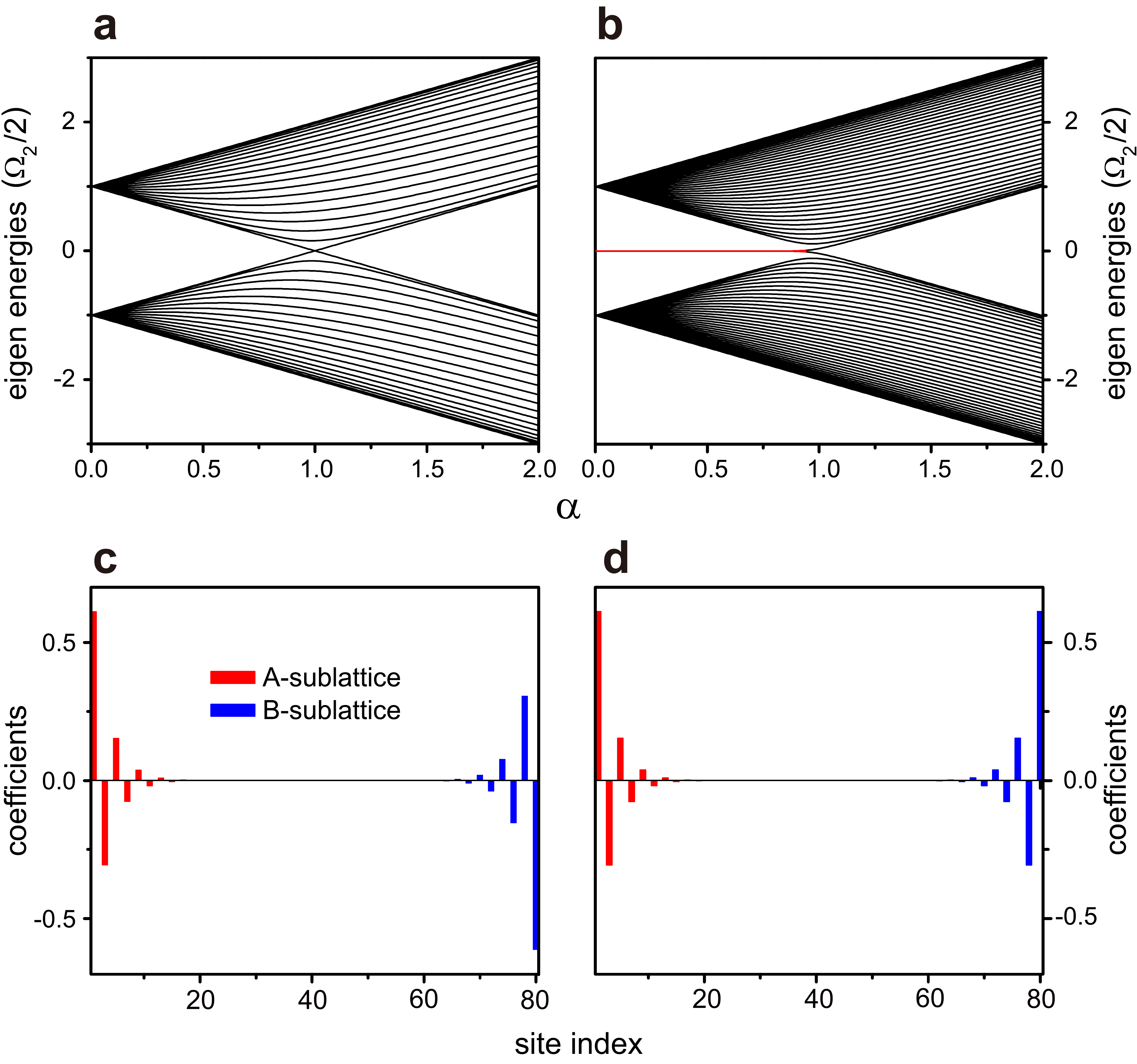}
\caption{The electronic structure of the SSH model with $N=40$ unit cells. The eigenenergies as functions
of the hopping amplitude ratio $\alpha=\Omega_1/\Omega_2$,  (a)  with and (b) without the periodic boundary condition.
The two red lines in (b) represent the appearance of the two edge states with almost-zero eigenenergies.
At $\alpha=0.5$, the coefficients of local wave functions, $\{|n, A\rangle, |n, B\rangle\}$,
for the two edge states are plotted in (c) and (d), respectively.  The columns in red and blue colors
correspond to the results of the $A$- and $B$-sublattices, respectively.
}
\label{fig:s3}
\end{figure}

The diagonalization of Eq.~(\ref{eq_S49}) gives rise to the band structures. The two eigenenergies of $H(\theta)$ are given by
\be
E_\pm(\theta) = \pm\frac{\hbar}{2}\sqrt{\Omega_1^2+\Omega_2^2+2\Omega_1\Omega_2\cos \theta},
\label{eq_S50}
\ee
where the $(+)$ and $(-)$ signs refer to the conduction and valence bands, respectively.
In Fig.~\ref{fig:s3}a, we plot the eigenenergies as functions of the hopping amplitude ratio, $\alpha=\Omega_1/\Omega_2$,
for the SSH model of $N=40$ unit cells with the periodic boundary condition.
Equation~(\ref{eq_S50}) shows that $E_\pm(\theta)$ is unchanged if
the intracell and intercell hopping amplitudes are swapped,
i.e., $E^\pr_{\pm}(\theta)=E_\pm(\theta)$ if $\Omega^\pr_1=\Omega_2$ and $\Omega^\pr_2=\Omega_1$.
The band gap is given by $\Delta_g=E_+(\theta=\pi)-E_-(\theta=\pi)=\hbar |\Omega_1-\Omega_2|$, which is closed under
the special condition of $\alpha=1$. %{\color{red}In addition to the dispersion relation (
To go beyond the dispersion relation, we inspect the corresponding Bloch eigenstates,
\be
|u_\pm(\theta)\rangle=\frac{1}{\sqrt{2}} \left[|\theta, A\rangle\pm e^{i\theta_\rmq} |\theta, B\rangle\right] =\frac{1}{\sqrt{2}} \left(\ba{c} 1 \\ \pm e^{i\theta_{\rmq}} \ea \right),
\label{eq_S51}
\ee
where $\theta_\rmq$ is a phase factor determined by
\be
\exp(i\theta_\rmq)=\frac{\Omega_1+e^{i\theta}\Omega_2}{\sqrt{\Omega_1^2+\Omega_2^2+2\Omega_1\Omega_2\cos \theta}}.
\label{eq_S52}
\ee
The Bloch eigenstates are not the same if $\Omega_1$ and $\Omega_2$ are swapped, and the local variation of $|u_\pm(\theta)\rangle$
 against the quasi-momentum $\theta$ (Berry connection) is sensitive to $\alpha$. Accordingly, we introduce an integrated quantity,
\be
\nu_\pm &=& \frac{1}{i\pi}\int_0^{2\pi} \langle u_\pm(\theta) |\partial_\theta |u_\pm(\theta)\rangle d\theta \no \\
&=& \frac{1}{2\pi}\int_0^{2\pi} \left(\frac{\partial \theta_\rmq}{\partial \theta}\right) d\theta,
\label{eq_S53}
\ee
which is a topological invariant:
$\nu_\pm = 1$ for $\alpha<1$ and $\nu_\pm = 0$ for $\alpha>1$. If we use the Bloch vector $\bm r_\pm(\theta)$
to represent $|u_\pm(\theta)\rangle$, Eq.~(\ref{eq_S53}) is rewritten as
\be
\nu_\pm = \frac{1}{2\pi} \int_0^{2\pi} \bm e_z\cdot\left[\bm r_\pm(\theta)\times d_\theta \bm r_\pm(\theta)\right] d\theta.
\label{eq_S54}
\ee
The topological invariant $\nu_\pm$ is thus equivalent to a winding number  of the curve
$\bm r_\pm(\theta)$ circulating around the center of the Bloch sphere.
Following the adiabatic theorem, the Bloch vector is proportional to the magnetic field,
i.e., $\bm r_\pm(\theta)=\pm \tilde{\bm B}_0(\theta)$ with  $\tilde{\bm B}_0(\theta)=\bm B_0(\theta)/|\bm B_0(\theta)|$.
Equation~(\ref{eq_S54}) is expressed by replacing $\bm r_\pm(\theta)$ with the reduced magnetic field $\tilde{\bm B}_0(\theta)$.
 In addition, the quantity $\gamma_\pm=\nu_\pm \pi$ is the Zak phase
of the Bloch eigenstate as the quasi-momentum traverses the 1D manifold of the FBZ~\cite{AtalaNatPhy13}.

To further understand the topology of the SSH model, we return to the original Hamiltonian in Eq.~(\ref{eq_S44}).
In Fig.~\ref{fig:s3}b, we plot the eigenenergies as functions of $\alpha$ for $N=40$ unit cells
without the periodic boundary condition. In the regime of $\alpha>1$, the SSH model is a conventional insulator.
The band gap, $\Delta_g\approx\hbar (\Omega_1-\Omega_2)$, is almost the same as
that from the bulk Hamiltonian in Eq.~(\ref{eq_S45}). In the opposite regime of $\alpha<1$,
two eigenstates, $|\Psi_\pm\rangle$, with almost-zero eigenenergies emerge in addition to the normal bulk eigenstates.
As shown by the example of $\alpha=0.5$ in Fig.~\ref{fig:s3}c,d, these two eigenstates
are mainly localized at the two ends of the SSH model. In particular, we can define two edge states,
$|\Psi_A\rangle$ and $|\Psi_B\rangle$, associated with the $A$- and $B$-sublattices respectively
(red and blue columns in Fig.~\ref{fig:s3}c,d).
The two eigenstates  are written as
\be
|\Psi_\pm\rangle = \frac{1}{\sqrt{2}} \left(|\Psi_A\rangle \pm |\Psi_B\rangle \right).
\label{eq_S55}
\ee
For a very large but finite $N$, the block Hamiltonian of the edge states
is given by
\be
H_{\rm edge} = \left(\ba{ll} \varepsilon & 0 \\ 0 & \varepsilon \ea \right),
\label{eq_S56}
\ee
in the basis set of $\{|\Psi_A\rangle, |\Psi_B\rangle\}$,
where $\varepsilon$ is the energy at the middle of the gap between the valence and conduction bands.
The two edge states, $|\Psi_A\rangle$ and $|\Psi_B\rangle$, become degenerate eigenstates, each at
one end of the SSH model.
In a finite-size SSH model, the significant difference of the band structures reveals
the separation of two topological phases: a trivial phase for $\alpha>1$ and a nontrivial phase for $\alpha<1$.

From the above calculations, we observe the separation of $\alpha<1$ and $\alpha>1$ for the topology of the
bulk Hamiltonian as well as the appearance of the edge states. In fact, these two behaviors are connected with each other:
The winding number is equal to the net number of the edge states at either end of the SSH model. This bulk-boundary
correspondence thus allows us to understand and predict the edge states using a topological invariant
of the bulk Bloch eigenstates.

\subsection{Quantum Simulation of the SSH Model}
\label{sec:s5b}

The SSH model is the simplest topological system, which can help us understand the topological behavior
of more complicated system in an intuitive way. The SSH model serves as a prototype system
in the quantum simulation of topological phases. One simulation approach is to create a 1D  lattice
in the real space~\cite{AtalaNatPhy13,JotzuNat14,AidelsburgerNatPhys15}.   %and measure a relevant topological invariant
Instead, the block diagonalization of the Hamiltonian
in the momentum space allows a mapping onto a spin-half particle driven by an external magnetic field.
The quantum simulation of the SSH model can thus be realized in a single qubit system. Based on the Hamiltonian
in Eq.~(\ref{eq_S49}) and the Bloch eigenstates in Eq.~(\ref{eq_S51}), we can prepare a
superposition state, $|u_\pm\rangle = (|0\rangle\pm|1\rangle)/\sqrt{2}$, and measure the relative Berry phase (Zak phase)
by rotating this quantum state along the equator of the Bloch sphere~\cite{ZZXPRA17}.

In an alternative approach, we can apply two
consecutive rotations, $U_z(-\pi/2)U_y(-\pi/2)$, to the Hamiltonian and the external magnetic field is
aligned in the $x$-$z$ plane. In a parameter sphere, the closed path of $0<\theta(t)<2\pi$
is transformed into a circular rotation along the longitudinal direction. In the spherical coordinate representation,
the magnetic field becomes a two-step function,
\be
\bm B_0(t) = \left\{ \ba{cll} (\Omega_2\sin\theta(t), 0, \Omega_1+\Omega_2\cos\theta(t))~~~&\mathrm{with}~&\phi=0~\mathrm{and}~\theta(t): 0\rightarrow \pi \\
(-\Omega_2\sin\theta(t), 0, \Omega_1+\Omega_2\cos\theta(t))~~~&\mathrm{with}~&\phi=\pi~\mathrm{and}~\theta(t): \pi\rightarrow 0 \ea \right..
\label{eq_S57}
\ee
where $\theta$ is the polar angle and $\phi$ is the azimuthal angle. The corresponding instantaneous spin-up eigenstate is
\be
|u_+(t)\rangle = \left\{ \ba{cll} \cos[\frac{\theta_\rmq(t)}{2}] |0\rangle + \sin[\frac{\theta_\rmq(t)}{2}] |1\rangle ~~~&\mathrm{with}~&\phi=0~\mathrm{and}~\theta(t): 0\rightarrow \pi \\
\cos[\frac{\theta_\rmq(t)}{2}] |0\rangle - \sin[\frac{\theta_\rmq(t)}{2}] |1\rangle ~~~&\mathrm{with}~&\phi=\pi~\mathrm{and}~\theta(t): \pi\rightarrow 0\ea \right. ,
\label{eq_S58}
\ee
with
\be
\theta_{\rm q}(t) = \arccos\left[ \frac{\Omega_1+\Omega_2\cos\theta(t))}{\sqrt{\Omega^2_1+\Omega^2_2+2\Omega_1\Omega_2\cos\theta}}\right] .
\label{eq_S59}
\ee
The instantaneous spin-down eigenstate can be similarly obtained.
Following the definition in Eq.~(\ref{eq_S53}), we calculate the Zak phase,
$\gamma_+ = -i \oint \langle u_+|\partial_\theta|u_+\rangle d\theta$, which depends on the
behavior of $\theta_{\rm q}$ when $\theta$ passes the singularity point, $\theta=\pi$.
To circumvent this difficult, we take the vector form in Eq.~(\ref{eq_S54}) and re-define
the winding number as
\be
\nu_+ = \frac{1}{2\pi} \oint\bm e_y\cdot\left[\bm r_+(\theta)\times d_\theta \bm r_+(\theta)\right] d\theta,
\label{eq_S60}
\ee
where the unit vector $\bm e_y$ along the $y$-direction is used due to the fact that  $\bm r_+$ is in the $x$-$z$ plane.
By transforming the spin-up state in Eq.~(\ref{eq_S58}) into the Bloch vector $\bm r_+$, we rewrite Eq.~(\ref{eq_S60})
as
\be
\nu_+ &=& \frac{1}{\pi} \int_0^\pi \left(\frac{\partial \theta_{\rm q}}{\partial \theta}\right) d\theta \no \\
&=& \frac{1}{\pi} \left[\theta_{\rm q}(\theta=\pi)-\theta_{\rm q}(\theta=0)\right].
\label{eq_S61}
\ee
Therefore, we can employ a half-circle rotation, $\theta(t): 0\rightarrow \pi$, to obtain the winding number.
In our experiment, we drive the spin to follow its instantaneous spin-up eigenstate. The adiabatic trajectory determines
the function of $\theta_{\rm q}(\theta)$, which then reveals the topology of the system.

%\subsection{Extension to the 2D Topological Systems}
%\label{sec:s5c}

The above methods of quantum simulation can be straightforwardly extended to the 2D topological systems. For a 2D FBZ,
the Bloch eigenstates of the two bands are functions of a 2D momentum, i.e., $|u_\pm(\bm k)\rangle$ with $\bm k=k_x \bm e_x+k_y \bm e_y$.
Here we introduce the Chern number using
\be
\calC h_\pm = \frac{i}{2\pi} \oint \nabla\times\langle u_\pm (\bm k)| \nabla | u_\pm (\bm k)\rangle\cdot d\bm S,
\label{eq_S62}
\ee
which can also be represented in a vector form as
\be
\calC h _\pm = \frac{1}{4\pi} \int \bm r_\pm(\bm k)\cdot\left(\frac{\partial \bm r_\pm(\bm k)}{\partial k_x}\times \frac{\partial \bm r_\pm(\bm k)}{\partial k_y} \right)dk_x d k_y.
\label{eq_S63}
\ee
The Chern number can be estimated from the integration of local Berry curvatures~\cite{HasanRMP10}.
The topology is fully determined if the whole structure of $|u_\pm(\bm k)\rangle$ or $\bm r_\pm(\bm k)$ is
extracted using the adiabatic trajectories.

%\begin{figure}
%\begin{center}
%\includegraphics[width=0.8\linewidth]{supp1.eps}
%\caption{\textbf{}}
%\end{center}
%\end{figure}